\documentclass[aps,pra,twocolumn,superscriptaddress,showpacs,amsmath,amssymb]{revtex4-1}
\usepackage{graphicx}

\usepackage[english]{babel}

\begin{document}


\title{Noise-induced topological transformations of vortex solitons in optical fibers filled with a cold atomic gas}


%
\author{A.V. Prokhorov}
\email{avprokhorov@vlsu.ru} \affiliation{Department of Physics and
Applied Mathematics, Stoletovs Vladimir State University, Gorky str.
87, Vladimir, 600000, Russia}
\author{M.G. Gladush}
\affiliation{Institute for Spectroscopy of the Russian Academy of
Sciences, Fizicheskaya Str. 5, Troitsk, Moscow, 142190, Russia}
\author{M.Yu. Gubin}
\affiliation{Department of Physics and Applied Mathematics,
Stoletovs Vladimir State University, Gorky str. 87, Vladimir,
600000, Russia}
\author{A.Yu. Leksin}
\affiliation{Department of Physics and Applied Mathematics,
Stoletovs Vladimir State University, Gorky str. 87, Vladimir,
600000, Russia}
\author{S.M. Arakelian}
\affiliation{Department of Physics and Applied Mathematics,
Stoletovs Vladimir State University, Gorky str. 87, Vladimir,
600000, Russia}
%
%
\date{\today}

\begin{abstract}
We consider the influence of optical and temperature-dependent atomic fluctuations on the formation and propagation of optical vortex solitons in dense media realized as hollow-core optical fibers filled with a cold atomic gas in presence of optical pumping. We show different perturbation-induced scenaria of complete destruction and smooth transformation of the topological characteristics of localized optical patterns in hollow-core fiber. The maximum levels of optical and atomic fluctuations at which the soliton regime can be maintained has been determined. The estimates for these levels show that it is possible to observe the optical vortex solitions in the core-filling gas of the fiber for temperatures smaller than the critical temperature of Bose-Einstein condensate.
\end{abstract}


\maketitle

\section{Introduction}
\label{intro} The investigation of the formation and propagation of
the localized spatial optical structures in different media
\cite{Kivshar} is of high interest to modern atomic physics. The importance of this topic is associated with a wide range of applications, such as data transmission and processing
\cite{Rozanov}. Among optical localized patterns, the greatest interest represents the special class of
optical topological structures known as optical vortices
\cite{PhysRevLett.105.213901}. The dark center of those structures could be
registered reliably in experiments even in the case of a strong
diffraction broadening of the optical beams
\cite{PhysRevLett.98.203601}. It is relatively easy to realize such
structures in laser cavities \cite{Smith:03} or spatially
inhomogeneous media \cite{Bezuhanov:04}. However, maintaining of
the stability conditions for such structures when they propagate in
the real media is challenging.  The soliton modes of optical
vortices have so far been observed quite rarely even within short
distances \cite{PhysRevLett.104.223902}. At the same time, the
distances for commercial telecommunications have not been
worked out at all. This can be explained by the fact that a full
scale solution to the problem of developing such information
channels requires a correct determination of the stability regions
\cite{JEPT:2012} under the experimental conditions,
as well as the stability tests with perturbations for the propagating
localized optical structures.

The formation of such stable optical structures occurs near the lasing threshold \cite{Rozanov} where the
nonlinear effects are very strong. Therefore the identification and analysis of the ranges of parameters at which the optical solitons are stable
requires a strict accounting of the higher-order nonlinearities and
nonlinear absorption effects in the system
\cite{Rozanov,PhysRevA.77.063810}. In order to experimentally realize an optical control over the nonlinear and dissipative parameters, it is the most beneficial to use the atomic
$\Lambda$--scheme of interaction between the localized structure of the probe and the pump fields. The
observations of the strong nonlinear effects in such a scheme can be
associated with the realization of the Raman regime
\cite{PhysRevA.77.063810,PhysRevA.39.3447} when the detunings of
the electromagnetic fields from the atomic transitions are much higher
than the corresponding decay rates (with the account of the optical depth of
the medium,  see \cite{PhysRevA.76.033805}).

Among promising media for observation and control of solitons,
we can mark out recently created hollow-core optical fibres
filled with a cold atomic gas \cite{PhysRevA.83.063830}. First, it would allow us to observe the coherent Raman scattering
and to study the competition between the nonlinear, diffraction and dissipative effects for the probe pulse over short distances \cite{Zheltikov,Kohler}.
Second, the use of cold atoms opens new possibilities of
precision and efficient control over the optical properties of the system. Indeed, in a cold medium we can neglect the effects of the spectral line broadening and splitting typical for hot atoms and solids.

It is a well known fact that the quality of transmission and the processing of information in
the optical data channel \cite{PhysRevLett.101.163901} is strongly
dependent on the efficiency of the nonlinear transformations, which
increases significantly with the increase of the density of the resonant atoms in
the medium. However, for the media with a large density of
optically active particles, it is necessary to take into account the
effective or the local value of the field acting on them
\cite{PhysRevA.29.2591,GladJETP:2011,PhysRevA.84.023828,PhysRevA.78.053827,Dolgaleva:12}.
The atom-field interaction in this case becomes more complicated. It is known in the
literature as the consequence of the near dipole-dipole
interactions (NDD). This may significantly modify the picture of the
phenomenon as the conventional Rabi frequency for an atomic
transition is comparable to the corresponding NDD correction term.
The estimations for ensembles of the resonant particles show that the
NDD or the local field effects may be very significant starting from the
density $\rho=10^{15}$ cm$^{-3}$. Due to this fact, a special attention should be paid to the problem
of obtaining stable dynamics of spatial solitons and optical
vortices in particular [3] in dense media with the additional accounting of perturbations in the medium and the field.

The influence of the noise effects on the soliton solutions of the nonlinear
Schr\"{o}dinger equation (NLS) has been well studied in the
literature, and a violation of the dynamic equilibrium in such a system
is typically observed for the values of the noise intensity
comparable to the average parameters of the problem
\cite{Mollenauer:96}. A particularly strong presence of the noise in
this problem manifests itself near the zero dispersion point where
the short-term noise-induced changes of the dispersion sign would quickly destroy the soliton.
The soliton condition in the such system can only be restored by an original pinning method \cite{Chertkov:02}. Analysis of the
fluctuations influence on the dissipative solitons is usually
performed artificially by introducing a noise source to the
Ginzburg-Landau equation (GLE)
\cite{PhysRevA.42.4661,PhysRevE.85.015205}. However, this approach
does not reveal the nature of the noise, and it is
therefore not possible to assess the contribution of various
physical processes into the development of
instabilities in the system. This is especially true in the case of
the optical solitons in gaseous media, where the density fluctuations at small scales can be comparable with the average values of density \cite{Einstein:1905,Smoluchowski:1906}. Even transition to the Bose-Einstein condensate (BEC) can not solve the problem completely because in such a system there are intricate dependencies of density fluctuations on the trapping geometry and the type of interaction between the particles \cite{Giorgini,Kocharovsky}.

In this paper, we consider the problem of obtaining a stable spatial
dynamics of optical vortices \cite{PhysRevLett.105.213901} in the
dense media of gas-filled hollow-core optical fibers in the presence of an optical pumping. The
problem is similar to the observation of temporal self-induced
transparency solitons in microstructured materials under the conditions of the
influence of the local field effects \cite{PhysRevA.84.023828}.
We also study the influence of random and/or
periodic perturbations of the system's parameters on the process of
stabilization and shape evolution during propagation of the vortex solitons.

This paper is structured as follows. In Section~\ref{sec:2} we present our formalism which contains the material equations derived from the Bogolyubov-Born-Green-Kirkwood-Yvon (BBGKY) hierarchy for reduced density matrices and correlation operators for a three-level atomic medium containing the local field effects in the interaction of individual impurity particles with electromagnetic fields. We study the temporal dynamics of the density matrix for the case of the $\Lambda$-configuration Raman interaction mode using optical vortices. The requirements for the atomic medium and the optical fields which allow a significant simplification of the problem are formulated. We as well analyze the spatial effects that arise in a self-consistent problem of the nonlinear scattering of the probe field in a dense medium of three-level atoms. We show the possibility to reduce the problem to the fifth-order nonlinear Ginzburg-Landau equation. In Section~\ref{sec:4} we use the variational methods and direct numerical simulation to determine the stability regions of spatial vortex solitons. In Section~\ref{sec:5} we describe the stability stress testing of optical vortices under perturbations of system's parameters. The main results and recommendations for the use of BEC-filled hollow fibers as information channels with dissipative vortex solitons is presented in Conclusion.

\section{Generalized density matrix equations for $\Lambda$-type atom--field interaction in a dense medium}
\label{sec:2}
\begin{figure}[b]
\includegraphics[width=\columnwidth]{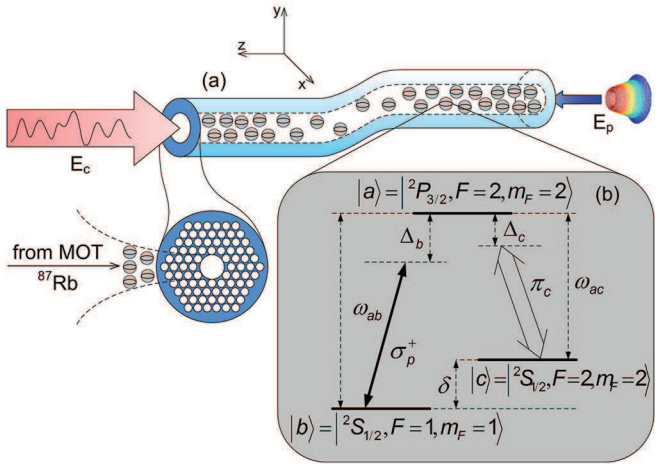}
\caption{\label{fig:1}(a) Model diagram of a hollow-core optical
fiber filled with an atomic gas (b) $\Lambda$-type interaction
configuration applied for $^{87}\rm Rb$ atoms, energy level
splitting between ${\left| c \right\rangle}$ and ${\left| b
\right\rangle}$ is $\delta =6.834\times 10^{9}$~s$^{-1}$, the dipole transition
matrix element for ${\left| a \right\rangle} \to {\left| b
\right\rangle}$ at wavelength $\lambda =780.241$~nm is
$\mu_{ab}=3.58\times 10^{-29}$~C$\cdot$m.}
\end{figure}

In this paper, we assume that a probe light beam of a given shape
${\bf E}_p$ with the center frequency $\omega_p$ propagates along
the $z$ axis of a hollow-core optical fiber filled with a gas of cold
atoms at a temperature $T$ near the critical temperature
\cite{Bagnato} of the phase transition:
\[T_{cr}=\frac{\hbar \omega_{0}}{k_{\textmd{B}}}\left(\frac{N}{\zeta(3)}\right)^{1/3},\]
where $\hbar$ is the Planck constant, $k_\textmd{B}$ is the Boltzmann constant, N is the total number of particles, $\omega_{0}$ is the trapping frequency, and $\zeta(3)$ is the zeta function of Riemann. It propagates in the direction opposite to cw pump radiation ${\bf E}_c$ as illustrated in Fig.~\ref{fig:1}a. In the Raman limit for the $\Lambda$-scheme of interaction the probe field detuning
$\Delta_b$ should be substantially greater than the relaxation rate
$\Gamma_{ab}$ ($\Gamma_{ac}$) from the excited state $|a\rangle$
(see Fig.~\ref{fig:1}b). This regime can be used for the formation of spatial optical solitons in the field of the probe radiation ${\bf E}_p$.
As applied to this problem, the optical depth of the medium $d_{0}$ is determined through the
characteristic linear dimension $a_{0}$ of topological structures
formed in the $XY$ plane as $d_{0}=g^{2} N a_{0}/(c\Gamma_{ac})$
(compare with \cite{PhysRevA.39.3447}). Here
$g=\mu_{ba}\sqrt{\omega/2\hbar\varepsilon_{0}V}$ is the atom-field
coupling constant, $\varepsilon=A_{p}\sqrt{2\varepsilon_0
V/\hbar\omega}$, $A_{p}$ is slowly varying amplitude of the probe
field, $V$ is the quantization volume, $N=\rho V_0$ is the number of
atoms in the interaction zone of volume $V_0$ for atomic density $\rho$, $c$ is the speed of light in vacuum,
$\varepsilon_0$ is the vacuum permittivity. The frequency separation
between levels $|b\rangle$ and $|c\rangle$ is
$\delta=6.834\times10^9$~s$^{-1}$, the dipole transition matrix
element for $|b\rangle\to|a\rangle$ is $\mu_{ba}=3.58\times10^{-29}$
C$\cdot$m.

Taking into account the large dipole moments of the desired transitions of $^{87}\rm Rb$ atoms, and the value of their density inside the hollow core fiber, one can conclude that the value of the NDD interactions
$\chi_{ba}=\rho\left|\mu_{ba}\right|^{2}/(3\hbar\varepsilon_0)$
which determines the effective Rabi frequency
$\Omega_{eff}=\Omega_{ba}+\chi_{ba}\sigma_{ba}$
\cite{PhysRevA.84.023828} can be close to the value of the Rabi
frequency of the probe field. Here $\sigma_{ba}$ is the
corresponding of the density matrix element. Thus, the nonlinear
optical effects arising in such a system will be determined, on
one hand, by the pump field parameters, and on the other hand, by
the number density of the atoms hosted by the hollow fiber.

This system is described by the following set of equations for the
density matrix elements:
\begin{widetext}
\begin{eqnarray}
 \label{eq:14}
 \nonumber
 \dot\sigma_{ba}&=&i\left(\Delta_b-\chi_{ba} \left(\sigma_{bb}-\sigma_{aa}\right)\right)\sigma_{ba}-ig\varepsilon \left(\sigma_{bb}-\sigma_{aa} \right)-i\left(\Omega +\chi _{ca} \sigma_{ca}\right)\sigma_{bc}-\frac{1}{2}\left(\Gamma_{ab} +\Gamma_{ac}\right)\sigma_{ba},\\
 \nonumber
 \dot\sigma_{ca}&=&i\left(\Delta_c-\chi_{ca}\left(\sigma_{cc}-\sigma_{aa}\right)\right)\sigma_{ca} -i\Omega \left(\sigma_{cc}-\sigma_{aa}\right)-i\left(g\varepsilon+\chi_{ba}\sigma_{ba} \right)\sigma_{cb}-\frac{1}{2}\left(\Gamma_{ab}+\Gamma_{ac}\right)\sigma_{ca},\\
 \nonumber
 \dot{\sigma }_{bc}&=&i\left(\Delta_b-\Delta_c \right)\sigma_{bc}+ig\varepsilon\sigma_{ac}-i\Omega^{*}\sigma_{ba}+i\chi_{ba}\sigma_{ba}\sigma_{ac} -i\chi_{ac}\sigma_{ac}\sigma_{ba},\\
 \nonumber
 \dot{\sigma }_{aa}&=&ig\varepsilon ^{*} \sigma _{ba} +i\Omega ^{*} \sigma _{ca} -ig\varepsilon \sigma _{ab} -i\Omega ^{*} \sigma _{ac} -\left(\Gamma _{ab} +\Gamma _{ac} \right)\sigma _{aa},\\
 \nonumber
 \dot\sigma_{bb}&=&ig\varepsilon \sigma_{ab}-ig\varepsilon^{*}\sigma_{ba}+\Gamma_{ab}\sigma_{aa}, \\
 \dot{\sigma}_{cc}&=&i\Omega\sigma_{ac}-i\Omega^{*} \sigma_{ca}+\Gamma_{ac} \sigma_{aa},
\end{eqnarray}
\end{widetext}
where $\Omega \equiv \Omega_{ca}$ and $g\varepsilon \equiv
\Omega_{ba}$ are the Rabi frequencies for the pump field and the
probe beam, respectively, and $\Delta_{b}$ and $\Delta_{c}$ are the
corresponding detunings.

Let us to consider applicability of the approximations for Eqs.~(\ref{eq:14}) under the condition of the specific requirements for our problem.
One of these requirements is
the presence of a significant polarization of the medium produced by
the contiguous transitions ${\left| a \right\rangle} \to {\left| b
\right\rangle}$ and ${\left| a \right\rangle} \to {\left| c
\right\rangle}$. It is necessary to provide nonlinear control over the
probe vortices via the pump wave. The key point here
is to choose the right balance between the Rabi frequency of the
pump field, its detuning from the resonance and the decay rate
of the excited state.

In general, the propagation equation for a probe field in a resonant
medium is
\begin{equation}
\label{eq:15} \left(\frac{\partial }{\partial t} +c\frac{\partial
}{\partial z} -ic\frac{D}{2} \nabla _{\bot }^{2} \right)\varepsilon
=-igN\sigma _{ba},
\end{equation}
where $\nabla_{\bot}^{2}=\partial^{2}/\partial x^{2}
+\partial^{2}/\partial y^{2}$, and $D=\lambda/\pi$ is the diffraction parameter in the
plane transverse to the axis $z$.

In order to solve the self-consistent problem of Eqs.~(\ref{eq:14})-(\ref{eq:15}) we assume, firstly,
that all atoms are initially at the level $b$, i.e., $\sigma_{bb}=1$,
$\sigma_{aa}=\sigma_{cc}=0$, and the population of the excited state
remains small throughout the interaction time, i.e., $\sigma_{bb}\cong
1$, $\sigma_{aa(cc)}\cong 0$ (while $\dot{\sigma }_{ii}\cong 0$
where $i=a,b,c$). Secondly, we believe that when
$\chi_{ca}=\chi_{ba}=\chi$, the contribution of the local field is
comparable to the probe Rabi frequency, but it can be
omitted for the pump transition, i.e.,
$g\varepsilon\ge\sigma_{ba}\chi_{ba}$ and
$\Omega\gg\sigma_{ca}\chi_{ca}$. As a result, the final system of
equations reduces to the following (simplified) form:
\begin{subequations}
\label{system:whole}
\begin{eqnarray}
\dot{\sigma}_{ba}&=&-\Gamma_{1} \sigma_{ba}-ig\varepsilon-i\Omega\sigma_{bc}-i\chi_{ba}\sigma_{ba},\label{system:1}\\
\dot{\sigma}_{ca}&=&-\Gamma_{2} \sigma_{ca}-ig\varepsilon \sigma_{cb}-i\chi_{ba} \sigma_{ba}\sigma_{cb}, \label{system:2}\\
\dot{\sigma}_{bc}&=&i\Delta_{3}\sigma_{bc}+ig\varepsilon\sigma_{ac}-i\Omega^{*}\sigma_{ba}, \label{system:3}\\
\dot{\sigma }_{bb}&=&ig\varepsilon
\sigma_{ab}-ig\varepsilon^{*}\sigma_{ba}, \label{system:4}
\end{eqnarray}
\end{subequations}
where $\Gamma_1=-i\Delta_b+1/2(\Gamma_{ab} +\Gamma_{ac})$,
$\Gamma_2=-i\Delta_c+1/2(\Gamma_{ab} +\Gamma_{ac})$, and
$\Delta_3=\Delta_b-\Delta_c$. In Eqs.~(\ref{system:whole}) the
contribution of the local field enters two ways. In
(\ref{system:1}), its action is trivial and produces an effective
frequency shift, but in (\ref{system:2}) it provides a more significant
effect, which is the appearance of the nonlinear coupling between the
atomic excitations (polarizations) in the probe $\sigma_{ba}$ and
magnetic $\sigma_{cb}$ transitions.
\begin{figure}[b]
\includegraphics[width=\columnwidth]{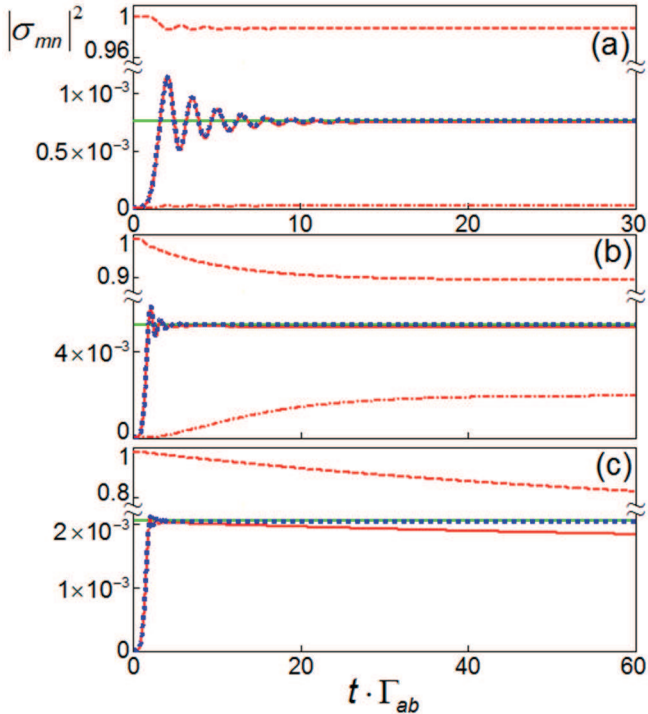}
\caption{\label{fig:2}Time dependence of the density matrix elements
in units of the excited state lifetime $\Gamma_{ab}^{-1}$ for (a)
the near-resonant and (b) Raman regimes of $\Lambda$--type interaction in
the limit of strong coupling; (c) for the near-resonant mode in the
limit of weak coupling. The thin solid line is for $|\sigma_{ba}|^{2}$,
dashed is for $|\sigma_{bb}|^{2}$ and the dash-dot line is for
$|\sigma_{cc}|^{2}$ are solutions of (\ref{eq:14}), the
thick dotted line is for $|\sigma_{ba}|^{2}$ obtained from
Eqs.~(\ref{system:whole}). The thick solid line corresponds to
$|\sigma_{ba}|^{2}$ which is calculated for the approximate
steady-state solution (\ref{eq:21}).}
\end{figure}

Figure~\ref{fig:2} shows the comparison of the
independent solutions of the exact system (\ref{eq:14}) and the
approximate (\ref{system:whole}) for different regimes of the
atom-field interactions at probe field with the switching time of
$\tau_{sw}=2\times 10^{-10}$~s and cw pump field. The atomic
and field parameters were chosen as follows: the characteristic size
of the optical beam $a_0=20$~$\mu$m, the atomic density
$\rho=1.01\times 10^{22}$ m$^{-3}$, the relaxation rate
$\Gamma_{ab}=\Gamma_{ac}=10^{9}$ s$^{-1}$, the probe field intensity
$I_p=0.22$ W$\cdot$cm$^{-2}$, the pump intensity $I_c=146.5$
W$\cdot$cm$^{-2}$, and the probe detuning $\Delta_b=-5\times 10^{9} $
s$^{-1}$. The corresponding Rabi frequencies can be calculated as
$\Omega=\mu_{ac}E_{c}/\hbar$ and $g\varepsilon=\mu_{ab}E_{p}/\hbar$
using the field strengths $E_{c(p)}=\sqrt{I_{c(p)}
2/c\varepsilon_0}$ and be $\Omega =1.13\times10^{10}$ s$^{-1}$ and
$g\varepsilon =4.4\times10^8$ s$^{-1}$, respectively
($g=1.3\times10^{6}$ s$^{-1}$ and $N=27.7\times10^{8}$). These
parameters meet the condition $\Gamma_{ab}<\Omega$ for the strong
atom-field coupling in the system.

At near-resonance interaction, when $\Delta_c/d_0\Gamma_{ab}\approx
0.1$ at $d_0=0.308$ and $\Delta_c = 3\times10^7$ s$^{-1}$, the
nonlinear coupling between the probe and pump fields is actually absent as it
follows from Fig.~\ref{fig:2}a. This is due to the fact that during
the interaction the ground state $| b \rangle$ is densely populated
($\sigma_{bb}\approx 1$) and atoms at levels
$| a \rangle$ and $| c \rangle$ are absent.
As a result, the polarizations on transitions $| a \rangle \to | b \rangle$ and $| a \rangle \to | c \rangle$ are weak.
This regime is typical for the linear EIT effect in a three-level medium
\cite{PhysRevA.65.022314}, or its non-linear analogue
\cite{PhysRevA.72.013804} at a significant increase of the pump
intensity. However, in this case, the solutions to the Eqs.~(\ref{eq:14}) and (\ref{system:whole}) for the matrix element
$\sigma_{ba}$ can approximate one another with sufficient accuracy
as a proof of validity of approximations used in the derivation of
(\ref{system:whole}).

Quite a different dynamics related to the nonlinear interaction of the
fields can be expected when we chose the Raman interaction mode
$\Delta_c>\Omega>d_0\Gamma_{ab}$. This interaction regime is
well-known \cite{PhysRevA.77.063810,PhysRevA.39.3447} and well
studied, but it did not get a strong ``scientific resonance'' in
terms of its prospects of practical use in modern optical
technologies, as it happened with EIT. This regime is characterized by
a set of curves for the matrix elements in Fig.~\ref{fig:2}b,
obtained for the same parameters as in Fig.~\ref{fig:2}a but with
a larger detuning $\Delta_c=3\times 10^{10}$ s$^{-1}$ and condition
$\Delta_c/d_{0}\Gamma_{ac}\approx 100$. In this case the atomic
state $| b \rangle$ is effectively depopulated giving the population
mostly to the state $| c \rangle$. This presupposes the emergence of a
significant polarization in transition $| a \rangle \to | b
\rangle$. It appears obvious if we compare the scale of values
$\sigma_{ba}$ in Figs.~\ref{fig:2}a and \ref{fig:2}b (the same is
true for the transition $| a \rangle \to | c \rangle $). This fact
leads to the establishment of a nonlinear energy transfer between
the modes of the probe and pump fields. This is the interaction regime we
are going to use further in our analysis.

The weak coupling in the atom-field system $\Gamma_{ab}>\Omega$ for
both regimes in Figs.~\ref{fig:2}a and \ref{fig:2}b produces a
significant discrepancy between the solutions obtained from
(\ref{eq:14}) and (\ref{system:whole}) as one may see in
Fig.~\ref{fig:2}c at $\Delta_c=3\times 10^{7}$ s$^{-1}$,
$\Omega=1.13\times 10^{8}$ s$^{-1}$. In particular, this is due to the
violation of the condition $\Omega\gg\sigma_{ca} \chi_{ca}$ and the
need to account for the effects of the local field for the transition $|
a \rangle \to | c \rangle$. This fact complicates the problem and
requires a direct numerical simulation of the self-consistent
problem (\ref{eq:15})-(\ref{system:whole})
\cite{PhysRevA.76.023827}. It makes further analysis of
opportunities to obtain solitons for this regime very
troublesome because of major transformations of the matrix elements.

Now let us find the steady state solution for the density matrix
element $\sigma_{ba}$ of the probe transition in a form that depends
only on the material parameters of the media and the
characteristics of the optical fields considering various cases of
atom-field interaction. In order to do this, we should solve
Eqs.~(\ref{system:whole}) in two steps. The first step is to define
the polarization of the system at the lower levels $\sigma_{cb}$ as
based on the approximation of constant populations and atomic
polarizations in the steady state, i.e.,
$\dot\sigma_{aa}=\dot\sigma_{bb}=\dot{\sigma}_{cc}=0$ and
$\dot\sigma_{ba}=\dot\sigma_{ca}=\dot\sigma_{bc}=0$. At this point
we obtain from Eqs.~(\ref{system:whole}) an algebraic equation for
the polarization $\sigma_{bc}$:
\begin{equation}
\Omega \chi g\varepsilon^{*} \sigma _{bc}^{2} +i\left(\Gamma _{1}
g^{2} \left|\varepsilon \right|^{2} +\Gamma_{2}^{*} A\right)\sigma
_{bc} +ig\varepsilon \Omega^{*} \Gamma _{2}^{*} =0,
\label{eq:17}
\end{equation}
the roots of which are the following:
\begin{equation}
\sigma _{bc} =\frac{-i\left(\Gamma _{1} g^{2} \left|\varepsilon
\right|^{2} +\Gamma _{2}^{*} A\right)\pm \sqrt{D} }{2\Omega \chi
g\varepsilon ^{*} }, \label{eq:18}
\end{equation}
where $D=-\left(\Gamma _{1} g^{2} \left|\varepsilon \right|^{2}
+\Gamma _{2}^{*} A\right)^{2} -4i\chi g^{2} \left|\Omega \right|^{2}
\Gamma _{2}^{*} \left|\varepsilon \right|^{2} $, $A=\left|\Omega
\right|^{2} -i\Delta _{3} \left(\Gamma _{1} +i\chi \right)$, $\chi
\equiv \chi _{ba} $.

Solutions (\ref{eq:18}) determine, in fact, the two branches of spin
excitations that occur at the transition between levels $| b
\rangle$ and $| c \rangle$ (see Fig.~\ref{fig:1}). The solution that
contains the minus sign in (\ref{eq:18}) leads to a problem with the
saturating nonlinearity $\sigma_{bc}\approx 1/{\varepsilon}$ and is
not considered in this paper. Expansion of the other solution, that
is positive, in series of the pump field $\varepsilon $ gives the
following relationship:
\begin{eqnarray}
\label{eq:19}
\sigma_{bc}&\approx& -\frac{g\Omega^{*}}{A} \varepsilon +\frac{g^{3}\Omega^{*}}{\Gamma_{2}^{*} A^{2}} \left(\Gamma_1+i\frac{\left|\Omega \right|^{2} \chi }{A} \right)\left|\varepsilon \right|^{2} \varepsilon \\
\nonumber\ &-&\frac{g^{5} \Omega ^{*} }{\left(\Gamma _{2}^{*}
\right)^{2} A^{3} } \left(\Gamma _{1}^{2} +\frac{3i\left|\Omega
\right|^{2} \chi \Gamma _{1} }{A} -\frac{2\chi ^{2} \left|\Omega
\right|^{4} }{A^{2} } \right)\left|\varepsilon \right|^{4}
\varepsilon.
\end{eqnarray}
Our second step is to use the equations for both $\dot\sigma_{ca}$
and $\dot\sigma_{bc}$ in (\ref{system:whole}) to write the atomic
polarization from the probe transition:
\begin{widetext}
\begin{equation}
\label{eq:20}
\sigma_{ba}=\frac{\left(\Delta_{3}+\displaystyle{i\frac{g^{2}
\left|\varepsilon \right|^{2} }{\Gamma _{2}^{*} }} \right)\sigma
_{bc} +\displaystyle{i\frac{g\varepsilon \chi}{\Omega \Gamma
_{2}^{*}} \left|\sigma_{bc} \right|^{2}}
\left(\Delta_3-i\frac{g^2\left|\varepsilon
\right|^2}{\Gamma_2}\right)} {\Omega ^{*}-\displaystyle{\frac{g^2
\left|\varepsilon\right|^2\chi^2}{\Omega \left|\Gamma _{2}
\right|^2}\left|\sigma_{bc}\right|^{2}}}
\end{equation}
\end{widetext}
Now we substitute expansion (\ref{eq:19}) into Eq.~(\ref{eq:20}) and
perform the secondary expansion to get an approximate solution
\begin{widetext}
\begin{eqnarray}
\nonumber \sigma_{ba}&\approx& \frac{g\Delta_3}{A}\varepsilon
-i\frac{g^3}{\Gamma_2^* A} \left(1+i\Delta_3\left(\frac{\Gamma_1}{A}+i\frac{\chi }{A^*} +i\frac{\chi \left|\Omega \right|^{2} }{A^2}\right)\right)\left|\varepsilon \right|^{2} \varepsilon \\
\nonumber &+&g^5\Biggl(\frac{1}{\left(A\Gamma_{2}^{*}
\right)^{2}}\left[i\left(1-\frac{\chi \Delta _{3} }{A^{*} }
\right)\left(\Gamma _{1} +i\frac{\chi \left|\Omega \right|^{2} }{A}
\right)-\frac{\Delta _{3} }{A}
\left(\Gamma_1^2-2\frac{\chi^2\left|\Omega\right|^4}{A^2}
+3i\frac{\chi\Gamma_1\left|\Omega \right|^2}{A}\right)\right]\\
&-&\frac{i\chi}{\left|A\right|^2\left|\Gamma_2\right|^2}
\left(i+\frac{\Delta _{3} }{A^{*} } \left\{\Gamma _{1}^{*}
-i\frac{\chi \left|\Omega \right|^{2} }{A^{*} } \right\}-i\frac{\chi
\Delta _{3} }{A} \right) \Biggr)\left|\varepsilon \right|^{4}
\varepsilon. \label{eq:21}
\end{eqnarray}
\end{widetext}
The solution (\ref{eq:21}), shown in Fig.~\ref{fig:2} with a thick solid line, is in agreement with the solutions
(\ref{eq:14}) and (\ref{system:whole}) to a high degree of accuracy. This serves as a proof of the correctness of (\ref{eq:21}). Let us note
that if solving Eqs.~(\ref{system:whole}) in the EIT limit,  when it
can be assumed that $\sigma_{ac}=0$, the expression for
$\sigma_{bc}$ contains only the first term on the right-hand side of
(\ref{eq:19}) (see \cite{PhysRevA.65.022314}). In its turn,
Eq.~(\ref{eq:20}) now becomes
\begin{equation}
\label{eq:22}
\sigma_{ba}=\frac{\Delta_3\sigma_{bc}}{\Omega^*}=\frac{g\Delta_3}{A}\varepsilon
\end{equation}
and determines the appearance of the phase modulation and the
absorption (amplification) of the probe pulse in the atomic medium.
We found that the appropriate solution for this linear mode (\ref{eq:22})
for $\sigma_{ba}$ with sufficient accuracy coincides with
the solutions of (\ref{eq:14}) and (\ref{system:whole}) in Fig.~\ref{fig:2}a.
The later proves the validity of the theory and the limiting cases as well.

For further analysis we will focus on the situation shown in
Fig.~\ref{fig:2}b and move on to the study of the spatial dynamics
of optical beams in the hollow-core fiber filled with resonant atoms
as shown in Fig.~\ref{fig:1}, and consider the possibility of obtaining a
special kind of stable structures known as vortex solitons
\cite{PhysRevLett.105.213901} of the probe field.

\section{Variational approach and numerical simulation for vortex solitons in a three-level medium}
\label{sec:4}
In the Raman limit for the $\Lambda$-type interaction
scheme (Fig.~\ref{fig:1}) after substituting (\ref{eq:21}) into the
propagation equation (\ref{eq:15}), the self-consistent problem of
spatial dynamics (\ref{eq:15})-(\ref{system:whole}) is reduced to
the well-known form of the Ginzburg-Landau equation (see
\cite{PhysRevLett.105.213901}):
\begin{eqnarray}
\label{eq:23} &&\left(\frac{1}{c}\frac{\partial}{\partial
t}+\frac{\partial }{\partial z}\right)\varepsilon
-i\frac{D}{2}\left(\frac{\partial^2 \varepsilon}{\partial x^2} +\frac{\partial^2 \varepsilon}{\partial y^2}\right)\\
\nonumber
&&-i\gamma_2\left|\varepsilon\right|^2\varepsilon+i\gamma_4\left|\varepsilon\right|^4\varepsilon
=-\alpha_I\varepsilon-\alpha_2\left|\varepsilon
\right|^2\varepsilon-\alpha_4\left|\varepsilon \right|^4\varepsilon,
\end{eqnarray}
with the following corresponding coefficients:
\[\gamma_{2}=\mathrm{Im}\left\{-\frac{g^{4} N}{A\Gamma_{2}^{*} c} \left(1+i\Delta_{3}
\left\{\frac{\Gamma_1}{A}
+\frac{i\chi}{A^*}+i\frac{\chi\left|\Omega\right|^2}{A^2}
\right\}\right)\right\}\] is the cubic nonlinearity;
\begin{eqnarray*}
\gamma_4&=&\mathrm{Im}\Biggl\{\frac{ig^{6} N}{c}
\Biggl(\frac{1}{\left(A\Gamma_2^*\right)^2}
\Biggl[i\left(1-\frac{\chi\Delta_3}{A^*}\right)
\left(\Gamma _{1} +i\frac{\chi \left|\Omega \right|^{2} }{A}\right)\\
&-&\frac{\Delta_3}{A}
\left(\Gamma_1^2-2\frac{\chi^2 \left|\Omega \right|^{4} }{A^{2} } +3i\frac{\chi \Gamma _{1} \left|\Omega \right|^{2} }{A} \right)\Biggr]\\
&-&\frac{i\chi}{\left|A\right|^2\left|\Gamma_2\right|^2}
\left(i+\frac{\Delta_3}{A^*}\left\{\Gamma_1^*-i\frac{\chi\left|\Omega\right|^2}{A^*}
\right\}-i\frac{\chi\Delta_3}{A}\right)\Biggr)\Biggr\}
\end{eqnarray*}
is the quintic nonlinearity;
\[\alpha_I=Im\left\{\frac{g^2 N\Delta_3}{Ac}\right\}\]
is the linear loss coefficient;
\[\alpha_{2} =\mathrm{Re}\left\{\frac{g^{4} N}{A\Gamma _{2}^{*} c} \left(1+i\Delta _{3} \left\{\frac{\Gamma _{1} }{A} +\frac{i\chi }{A^{*} } +i\frac{\chi \left|\Omega \right|^{2} }{A^{2} } \right\}\right)\right\}\]
is the cubic loss;
\begin{eqnarray*}
\alpha_{4}
&=&\mathrm{Re}\Biggl\{\frac{ig^{6}N}{c}\Biggl(\frac{1}{\left(A\Gamma_2^*\right)^2}
\Biggl[i\left(1-\frac{\chi \Delta_3}{A^*}\right)
\left(\Gamma_{1}+i\frac{\chi \left|\Omega \right|^{2} }{A} \right)\\
&-&\frac{\Delta_3}{A}
\left(\Gamma_1^2-2\frac{\chi^2\left|\Omega\right|^{4}}{A^{2}}+3i\frac{\chi\Gamma_{1} \left|\Omega \right|^{2}}{A} \right)\Biggr]\\
&-&\frac{i\chi}{\left|A\right|^{2}\left|\Gamma_{2}\right|^{2}}
\left(i+\frac{\Delta_{3}}{A^{*}}
\left\{\Gamma_1^*-i\frac{\chi\left|\Omega
\right|^2}{A^*}\right\}-i\frac{\chi\Delta_3}{A}
\right)\Biggr)\Biggr\}
\end{eqnarray*}
is the quintic loss.

It is clear that Eq.~(\ref{eq:23}) is the result of a complicated
nonlinear interaction between the probe field and the pump field
which arises solely due to the presence of significant values of
polarizations of the optical transitions in the
$\Lambda$-configuration \cite{PhysRevA.76.033805}. Even though the
locally acting field used in this theory does not produce new terms
in Eq.~(\ref{eq:23}) it, however, gives certain corrections to the
picture of the nonlinear interaction (terms with $\chi$--factor)
for the case of an optically dense medium. Let us note that in
another limit of the $\Lambda$-configuration the near-resonance
conditions, i.e. $\Delta_b\ll\Gamma_{ab}$, would simplify
Eq.~(\ref{eq:23}) by eliminating its nonlinear terms.  The local
field in this case would be insignificant as it can only lead to an
additional phase modulation in the probe field
\cite{PhysRevA.82.013815}.

In order to start the analysis of (\ref{eq:23}) we will first
transform to the moving coordinate system $T=t-z/c$ and perform the
change of variables
$u=\varepsilon/\sqrt{\left|\varepsilon_{in}\right|^2}$, $\xi
=z/L_{df}$, $X=x/a_{0}$, $Y=y/a_{0}$. Our second step is to define
the following main characteristic lengths: $L_{\gamma
2}=1/(\gamma_2\left|\varepsilon_{in}\right|^{2})$ and $L_{\gamma
4}=1/(\gamma_4 \left|\varepsilon_{in} \right|^4)$ are the
nonlinearities of the third and the fifth orders respectively;
$L_{\alpha I}=1/\alpha_I$ describes the linear losses, $L_{\alpha
2}=1/(\alpha _{2} \left|\varepsilon _{in} \right|^{2} )$ and
$L_{\alpha 4}=1/(\alpha_4 \left|\varepsilon_{in}\right|^4)$ describe
the nonlinear losses of the third and fifth orders; $L_{df}
=a_{0}^{2}/D$ is the diffraction length, where $\varepsilon_{in}$ is
the reduced amplitude of the beam at the entrance point. After multiplying both
sides of (\ref{eq:23}) by $L_{df}$ we finally get
\begin{equation}
i\frac{\partial U}{\partial \xi } +\frac{1}{2}\left(\frac{\partial
^{2} U}{\partial X^2} +\frac{\partial^2 U}{\partial Y^2}
\right)+\left|U\right|^2 U-\nu \left|U\right|^{4} U=Q,
\label{eq:24}
\end{equation}
where $Q=i\bigl[-\delta U-\phi \left|U\right|^{2} U-\mu
\left|U\right|^{4} U\bigl]$ is the dissipative part, for which we
have the following notations $U=uN$ and $N^{2}=L_{df}/L_{\gamma2}$,
and the main characteristic parameters: $\delta =L_{df}/L_{\alpha
I}$, $\phi =L_{\gamma 2}/L_{\alpha 2}$, $\mu=L_{\gamma
2}^2/(L_{\alpha 4}L_{df})$, $\nu=L_{\gamma 2}^2/(L_{\gamma 4}
L_{df})$.
\begin{figure}[t]
\includegraphics[width=\columnwidth]{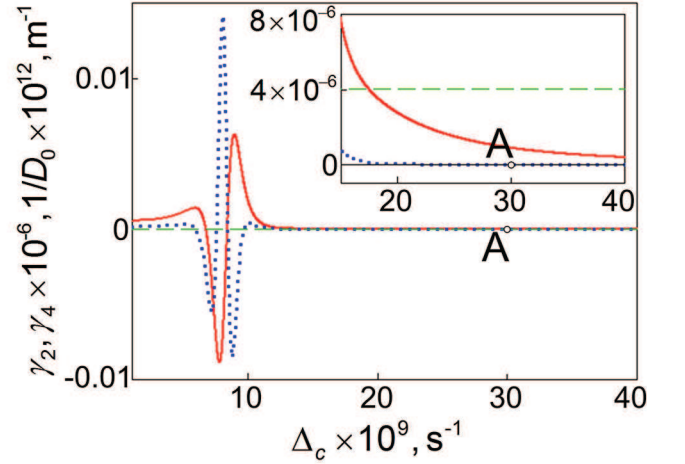}
\caption{\label{fig:3}Frequency dependences of the third $\gamma_2$
(solid line) and fifth-order $\gamma_4$ (dotted line)
nonlinearities, and the backward diffraction (dashed line).
The inset shows the enlarged image of the region near the point $A$.
The interaction parameters are the same as in Fig.~\ref{fig:2}(b).}
\end{figure}

\begin{figure}[t]
\includegraphics[width=\columnwidth]{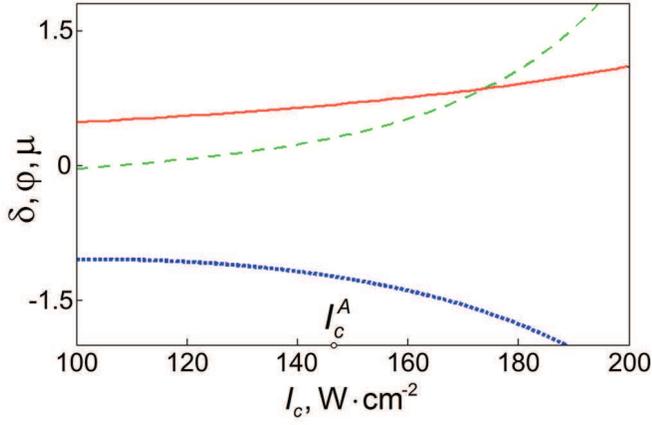}
\caption{\label{fig:4}Parameters of the dissipative part of the
non-dimensional Ginzburg-Landau equation $\delta$ (solid line),
$\phi$ (dotted line), and $\mu$ (dashed line) as functions of the
pump field intensity. The interaction parameters are the same as for
point A in Fig.~\ref{fig:3}.}
\end{figure}

For the formation of a conservative optical soliton the
interaction parameters should be selected in such a way that the
characteristic nonlinear and diffraction lengths were approximately
comparable or multiples of each other \cite{Agrawal:book:2001}. This
can be achieved near the region of strong transformations of the
nonlinear  coefficients by varying the detuning of the pump field as
it is shown in Fig.~\ref{fig:3}. In particular, for the point
$A$ with detuning $\Delta_{c}^{A}$ the diffraction length should be only $L_{df}=1.61$~mm so for the positive nonlinearity, i.e. $\gamma_2>0$, we may
expect focusing and formation of a stable spatial soliton. However,
following the concept of dissipative solitons, to maintain the
energy of bright solitons it would require interchanging between
absorption and gain effects in different areas of the special
envelope in cross-section of probe beam. In particular, these conditions can be realize when necessary inequalities $\delta >0$, $\phi
<0$, $\mu >0$ are to be met \cite{Lect}. Figure~\ref{fig:4} shows
the typical dissipative parameters as functions of the pump field
intensity $I_c$ at a fixed detuning $\Delta_c^A$ as in
Fig.~\ref{fig:3} near the gain threshold. Under conditions of
differently directed variation of the dissipative coefficients with increasing
pump intensity we can select $I_c$ for fixed $\Delta_c^A$ (or vice
versa) to satisfy the condition of dynamic equilibrium
for nonlinear and dissipative processes in the system.

Using the variational approach for the analysis of (\ref{eq:24}) we may
determine the ranges of parameters for which stable spatial solitons
can occur. Our particular interest is focused on an important class
of dissipative spatial vortex solitons of the kind described in
\cite{PhysRevLett.105.213901}:
\begin{eqnarray}
\nonumber
U&=&A_0 A\left(\frac{r}{R_0 R}\right)^S \exp{\Biggl\{-\frac{r^2}{2\left(R_0 R\right)^2}}\\
 &+&i\left(C\frac{r^2}{R_0^2}+S\theta+\Psi\right)\Biggr\},
\label{eq:25}
\end{eqnarray}
where $r=\sqrt{X^2+Y^2}$, $\theta$ is the angle in spherical
coordinates and $A$, $R$, $C$, and $\Psi$ are  the amplitude, the
spatial width, the curvature of the wave front and the phase of the
soliton, respectively. Parameter $S$ determines the topological charge
of the vortex soliton. In the special case of $S=1$ the
normalization coefficients $A_0$ and $R_0$ can be expressed in terms
of the full power normalization
\[P=\int_0^{2\pi}\int_0^\infty\left|U(r,\theta )\right|^{2} rdrd\theta =\left(\pi S! A_{0}^{2} R_{0}^{2} \right)A^{2} R^{2}.\]
In the simplest case, when $P=A^{2}R^{2}$, they are related as
$A_{0}=1/(R_{0}\sqrt{\pi})$ and $R_{0}=1$.

For a self-maintained special crater-shaped vortex solitons
(\ref{eq:25}) the system would require not only a balance of the
nonlinear, diffraction and dissipation effects \cite{Lect} but also
some presence of the optical diffusion \cite{PhysRevA.76.045803} or
a complicated type of modulation of either the refractive index
\cite{PhysRevA.82.023813} or the absorption coefficient
\cite{PhysRevLett.105.213901} or both. In particular, the static
parameter of the linear absorption can be replaced with a
space-dependent effective parameter $\delta_{\rm eff}=\delta-Vr^2$.
It would describe introduction of an additional saturable
absorber into the fiber filled with a cold gas. This is the case we
consider further in this paper and assume $V=-0.03$ everywhere
below. A detailed analysis of the stability of vortex solitons in
the linearization of the master equation written in a form more
general than (\ref{eq:23}) and with the optical diffusion
taken into account is given in \cite{Fedorov:2003}.

Now we use the Euler-Lagrange equation for (\ref{eq:24})--(\ref{eq:25}) and we get the system of equations for the variable parameters:
\begin{subequations}
\label{system2:whole}
\begin{eqnarray}
\nonumber
\frac{dA}{d\xi}&=&-\frac{5\phi A^{3} }{16\pi }-\frac{8\mu A^{5} }{81\pi ^{2} } \\
&&+A\left(-\delta -C+VR^{2} \right), \\
\frac{dR}{d\xi}&=&\frac{\phi A^{2} R}{16\pi } +\frac{2\mu A^{4} R}{81\pi ^{2} } +CR+VR^{3}, \\
\frac{dC}{d\xi}&=&-C^{2} +\frac{1}{8R^{4} } -\frac{A^{2} }{16\pi R^{2} } +\frac{2\nu A^{4} }{81\pi ^2 R^2}, \\
\frac{d\Psi}{d\xi}&=&\frac{3A^2}{8\pi}-\frac{10\nu
A^4}{81\pi^2}-\frac{1}{2R^2}.
\end{eqnarray}
\end{subequations}
For a low frequency modulation ($C^{2}\cong 0$), as in
\cite{PhysRevLett.105.213901}, we come to the following system of
equations, which simplifies the search for the fixed points for
(\ref{system2:whole}):
\begin{subequations}
\label{system3:whole}
\begin{eqnarray}
\nonumber
&&C=\frac{A^{2} \left(-81\pi \phi -32A^{2} \mu \right)}{1296\pi^2}\\
&&\;\;\;\;\;\;\;\;\;\;\;\;\;\;\;\;\;\;\;\;\;\;\;\;
-\frac{162\pi ^{2} V}{A^{2} \left(81\pi -32A^{2} \nu \right)}, \\
&&R^{2}=\frac{162\pi ^{2} }{A^{2} \left(81\pi -32A^{2} \nu \right)}, \\
&&-\frac{5\phi A^3}{16\pi}-\frac{8\mu
A^5}{81\pi^2}+A\left(-\delta-C+VR^2\right)=0.
\end{eqnarray}
\end{subequations}
Equations~(\ref{system3:whole}) give $16$ steady state roots, among which only
two correspond to the physical conditions on the energy and the width of
the vortex soliton ($A>0,R>0$ and $A,R\in\Re$). Besides, only one
of them would have the maximum value of A and negative $C$ and be stable \cite{PhysRevLett.105.213901}.

\begin{figure}[t]
\includegraphics[width=\columnwidth]{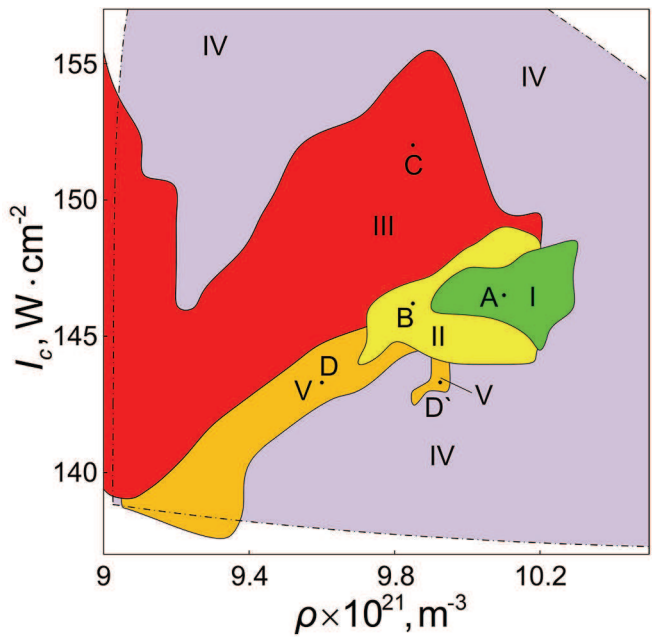}
\caption{\label{fig:5}Parametric plane (the intensity of the pump
field $I_c$ vs the dopant density $\rho$). The dash-dot line margins
the area of stability for an axisymmetric vortex soliton obtained by
the variational approach. The allocated areas are obtained by
direct numerical simulations and correspond to: I - axisymmetric
vortex solitons, II - single-hump vortex solitons, III -
fundamental solitons, V - nonstationary localized structures,
IV - unstable structures. The interaction parameters correspond to
Fig.~\ref{fig:4}.}
\end{figure}

Figure~\ref{fig:5} shows the parametric plane formed by the
following parameters: the density of resonant atoms in the system
$\rho$ and the intensity of the pump field $I_c$. The region bounded
by the dash-dot line refers to stability of axisymmetric vortex
solitons and arise for the selected physical solution of
(\ref{system3:whole}) \cite{PhysRevLett.105.213901}. This area was
determined from the analysis of the eigenvalues of the Jacobi matrix
for Eqs.~(\ref{system2:whole}), i.e., for condition
$Re\left(\lambda_j\right)<0$, where $j=1,2,3$ \cite{Diff}, and
corresponds to the point of a stable focus.

Direct numerical simulations of Eq.~(\ref{eq:24})
taking into account the initial angular perturbations for $R$ and $C$
\cite{PhysRevB.81.035202} in (\ref{eq:25}) show that the true stability area for axisymmetric vortex
soliton I
(in Fig.~\ref{fig:5}) is much smaller. The
region of stability obtained using the variational approach
appears to have a ``fine'' structure in the form of separate zones
for solitons with modified shapes.

\begin{figure}[t]
\includegraphics[width=0.807\columnwidth]{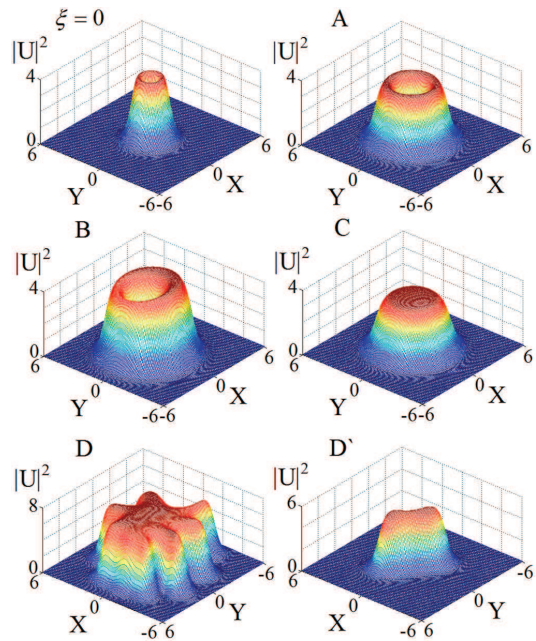}
\caption{\label{fig:7}Results of direct numerical simulations of
Eq.~(\ref{eq:24}), i.e. spatial shapes in $(X,Y)$-plane of optical
beams at the entrance $\xi=0$ and after propagation the distance of
$\xi=100000$ in the gas-filled fiber in the presence of azimuthal
perturbations. The letter in the upper left corner of each image
corresponds to a point in the parameter plane in Fig.~\ref{fig:5},
the coordinates of which were used to calculate the parameters of
the Eq.~(\ref{eq:24}).}
\end{figure}

In particular, in the area II in Fig.~\ref{fig:5} (parameters as
in Eq.~(\ref{eq:24}): $\nu=0.1653$, $\delta=0.6819$, $\phi
=-1.1998$, $\mu =0.2432$ at $\rho =9.85\times10^{21}$ m$^{-3}$ and
$I_c=146.2$ W$\cdot$cm$^{-2}$ for point B) we find a spontaneous
transition of an axisymmetric vortex into a single-humped asymmetric
stable vortex soliton, presented in image B in Fig.~\ref{fig:7}.
Bifurcations of this type are described in
\cite{PhysRevLett.105.213901}. Region III in
Fig.~\ref{fig:5} (parameters as in Eq.~(\ref{eq:24}): $\nu=0.2519$,
$\delta=0.7158$, $\phi=-1.2576$, $\mu=0.3169$ at $\rho=9.85\times
10^{21}$ m$^{-3}$, $I_c= 152$ W$\cdot$cm$^{-2}$ for point C) is
characterized by a high sensitivity of the system to initial angular
perturbations. During the propagation the vortex
soliton spontaneously loses its topological charge and transforms
into a new vortex-free stable state with $S=0$ (see image C in
Fig.~\ref{fig:7}).

In area V the account for angular perturbation in
Eq.~(\ref{eq:25}) leads to destruction of vortex solitons and
emergence of nonstationary localized structures in their place,
which, however, do not damp in time and demonstrate some ongoing evolution
\cite{Rozanov_OS:2003}. At point D in
Fig.~\ref{fig:5} the parameters in
Eq.~(\ref{eq:24}) take values $\nu=0.1043$, $\delta=0.674$,
$\phi=-1.14$, $\mu=0.1551$ at $\rho=9.6\times 10^{21}$ m$^{-3}$ and
$I_c=143.3$ W$\cdot$cm$^{-2}$. Here the axisymmetric vortex soliton
transforms in a multi-hump structure rotating during its propagation as in image D in
Fig.~\ref{fig:7}. It is also possible to observe a specific
double-hump optical structures \cite{Fedorov:2003} as in image D' in
Fig.~\ref{fig:7}. This structure occurs when $\nu=0.1297$,
$\delta=0.6633$, $\phi=-1.1848$, $\mu=0.228$ at $\rho=9.925\times
10^{21}$ m$^{-3}$ and $I_{c}=143.3$ W$\cdot$cm$^{-2}$. In area IV we
notice the loss of stability and complete damping of optical
solitons.

Let us note that point A (with the same set of parameters as in
Fig.~\ref{fig:3}) is indeed found in the area of stability,
calculated using both the variational approach and direct
simulations, which is in agreement with the qualitative analysis of
Eq.~(\ref{eq:24}). Thus, formation of vortex solitons in our medium with conditions for point A in Fig.~\ref{fig:5}
would require the initial spatial width and the curvature of the wave
front of the optical soliton in the form of Eq.~(\ref{eq:25}) be
$r_{R}=19$~$\mu$m and $C_R=-3.7\times10^{6}$~m$^{-2}$ respectively.
The dimensionless parameters are, therefore, $A=5.75$, $R=0.95$, and
$C=-0.0015$. The values calculated for Eq.~(\ref{eq:24}) for this regime
are $\nu=0.1899$, $\delta=0.675$,
$\phi=-1.2375$, and $\mu=0.3075$. The stabilization of optical vortices
is achieved within the characteristic length of $L_{ST}=4$~cm while
its rotation period is
$T_V=1.45$~cm (see image A in Fig.~\ref{fig:7}).

\begin{figure}[!]
\includegraphics[width=\columnwidth]{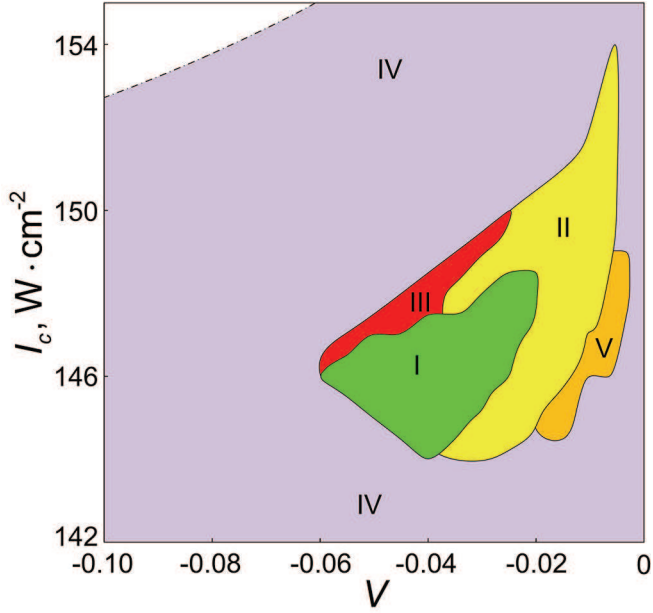}
\caption{\label{fig:6}Parametric plane (the intensity of the pump
field $I_c$ vs. the optical trapping parameter $V$). Interaction
parameters and numbering of areas are the same as in Fig.~\ref{fig:5}.}
\end{figure}

Figure~\ref{fig:6} shows the region of stability for vortex solitons
in the plane ($I_c,V$) at $\rho=1.015\times10^{22}$~m$^{-3}$. The
area of stability obtained by the variational approach extends to both
the positive and negative values of parameter $V$. The direct numerical
simulation of Eq.~(\ref{eq:24}) gives a narrower area and demonstrates that stable
dissipative vortices can exist only for negative values of $V$.

The fundamental point in this problem is consideration of the
local field effects. Indeed, if $\chi=0$, the area of stability is not simply
transformed, but it completely transcends the parametric plane shown
in Fig.~\ref{fig:4}. In this case, all the solutions obtained for
the optical solitons on the parametric planes in Figs.~\ref{fig:5},~\ref{fig:6} become unstable.

Further analysis and exploration of the other stability areas for
the vortex solitons require solution of the full nonlinear system
(\ref{system2:whole}) as well as an expanded multi-dimensional
numerical experiment within the domain of the system's parameters.

\section{Stability stress testing of optical vortices under perturbations of system's parameters}
\label{sec:5}
\begin{figure}[!]
\includegraphics[width=0.905\columnwidth]{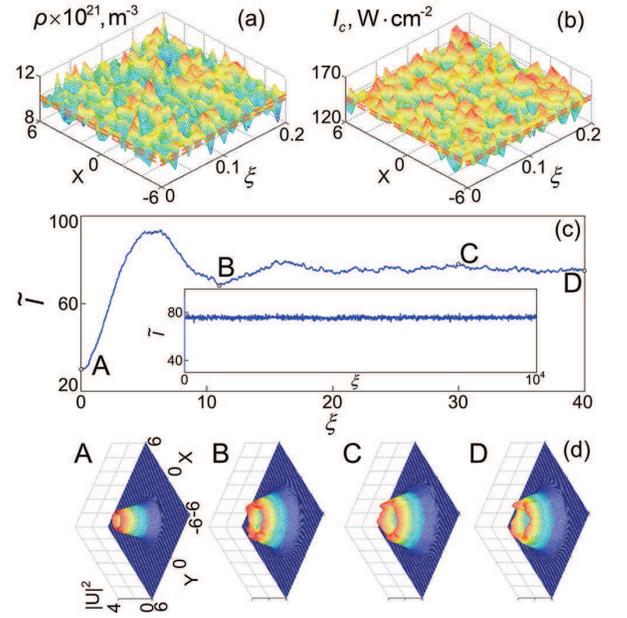}
\caption{\label{fig:8}The model of the statistical stability for the
axisymmetric vortex solitons with fluctuations of  the atomic
density $\rho$ (a) and the intensity of the pump field $I_c$ (b). The
range of the fluctuating quantities between the dashed lines
corresponds to the stability area in Fig.~\ref{fig:5}. The temporal
dynamics of the reduced power $\tilde{I}$ of the soliton is shown in
image (c), the shapes of the probe optical beam in image (d). The letter
at the upper left corner of each image (d) corresponds to the timepoint in
image (c).}
\end{figure}

Let us consider the evolution of an optical vortex
soliton (\ref{eq:25}) in terms of noise fluctuations or/and
additional modulations of the optical fields and core-filling gas in the fiber. Such perturbations may be
associated with variations of the pump field intensity $I_c
(X,Y,\xi)=I_{c}^{\rm det}+\zeta_I(X,Y,\xi)$ and the atomic
density $\rho(X,Y,\xi)=\rho^{\rm det}+\zeta_\rho(X,Y,\xi)$,
where $I_{c}^{\rm det}$ and $\rho^{\rm det}$ describe the
deterministic part of control parameters. The perturbative parameters $\zeta_I(X,Y,\xi)$ and
$\zeta_{\rho}(X,Y,\xi)$ are presented by three-dimensional matrices
of fluctuations or periodic spatial modulation of the corresponding
functions. Random perturbations of the intensity $\zeta_{I}^{\rm
rnd}(X,Y,\xi)$ we describe as the Gaussian white noise. The
density fluctuations in a Bose gas below the critical
temperature at transformation to the dimensional parameters have a
spatial correlation function in the form
\begin{equation}
\left\langle\zeta_{\rho}^{\rm rnd}(r_1)\zeta_{\rho}^{\rm
rnd}(r_2)\right\rangle =\rho_0(\delta(r_1-r_2)+\nu_n(r_1-r_2)),
\label{eq:28}
\end{equation}
where
\begin{eqnarray*}
r_{i}&=&(z_i,x_i,y_i),\\
\nu_{n}(r_{\rm rnd})&=&\frac{m_a k_\textmd{B} T\rho_b}{\pi\rho_0 \hbar^2
r_{\rm rnd}}+\frac{3m_{0}^{2} k_{\textmd{B}}^2 T^2}{4\pi^2\rho_0\hbar^4
r_{\rm rnd}^2}.
\end{eqnarray*}
In our numerical simulations we used the three characteristic
spatial scales, i.e., the sampling interval (simulation step) of $\xi_{st}=0.0011$, the
characteristic length of density variation (physical step) of
$\xi_{\rm rnd}=0.0068$ (equivalent to $r_{\rm rnd}=11$ $\mu$m) and
$\xi_{\rm var}=0.07$ for the characteristic distance at which the
mean value of the noise becomes zero. The selected physical step is so
great that $\nu_n(r_{\rm rnd})/\rho_0\approx 0.4\% $, so the second
term in the right-hand side of Eq.~(\ref{eq:28}) is neglected and we
can model the density fluctuations using the white noise and
spatial splines (Fig.~\ref{fig:8}a,b) for the convenience of
calculations at the intermediate points.

The periodic modulation of the pump field intensity
$\zeta_{I}^{\rm reg}(\xi )$ and the atomic density
$\zeta_{\rho}^{\rm reg}(\xi )$ with the amplitudes $\zeta_{I_0}^{\rm
reg}$ and $\zeta_{\rho_0}^{\rm reg}$ are assumed to be effective
only along axis $z$: $\zeta_{I}^{\rm reg}(\xi)=\zeta_{I_0}^{\rm
reg}\sin(2\pi\xi/L_I+\varphi_I)$ and $\zeta_{\rho}^{\rm reg}
(\xi)=\zeta_{\rho_0}^{\rm reg}\sin(2\pi\xi/L_{\rho}+\varphi_{\rho})$
with the specified spatial periods $L_{I}$, $L_{\rho}$ and random
initial phases $\varphi_{I(\rho)}$. The pulsations of the pump
intensity can have two sources: the artificial modulation at the
output of the pump laser or the intensity oscillations occuring in
the pump depletion and gain saturation \cite{PhysRevA.4.1175}. The
periodic modulation of the density can be caused by an external
mechanical action entailing excitation of acoustic waves in the
system.

The specificity of our problem lies in the possibility of studying
the generalized picture of the development of perturbations of each GLE
parameter and their influence on the dynamics of the soliton via the perturbations of the control parameters for
media and the fields. Quantitatively, this may be
reflected by the values of the noise strength $D_{n}=z_{\rm
var}(\sigma_n)^2$ for the GLE coefficients
$n=\left(\gamma_{2},\gamma_{4},\alpha_I,\alpha_2,\alpha_4\right)$ with their
mean-square deviations $\sigma_n$, calculated on the presence of random
perturbations for control parameters with the relative deviations $\varepsilon_I=\sigma_I/I_c^{\rm det}$
and $\varepsilon_{\rho}=\sigma_{\rho}/\rho^{\rm det}$. Here
$\sigma_I$ and $\sigma_{\rho}$ are the mean-square deviations for $I_c$ and $\rho$, respectively; $z_{\rm var}=\xi_{\rm var} L_{df}$. The possibility
in our numerical simulation of summarized accounting of noise +
periodic modulation of the system parameters completes
the full-scale stress testing picture for special optical structures which propagate in the hollow-core fiber.

\begin{figure}[!]
\includegraphics[width=\columnwidth]{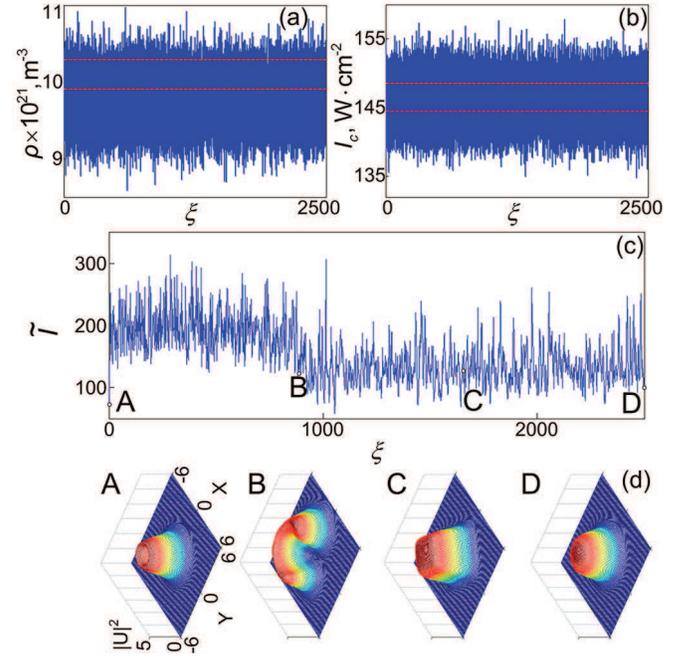}
\caption{\label{fig:9}The model of the ``breathing'' state as a
result of migration of the soliton solutions between regions III and
V in Fig.~\ref{fig:5} in  presence of fluctuations (a) the atomic
density $\rho$ and (b) the pump intensity $I_c$. Image (c)
shows the temporal dynamics of $\tilde{I}$; the probe beam shapes corresponding to
different timepoints from (c) are presented in (d).}
\end{figure}

\begin{figure}[!]
\includegraphics[width=\columnwidth]{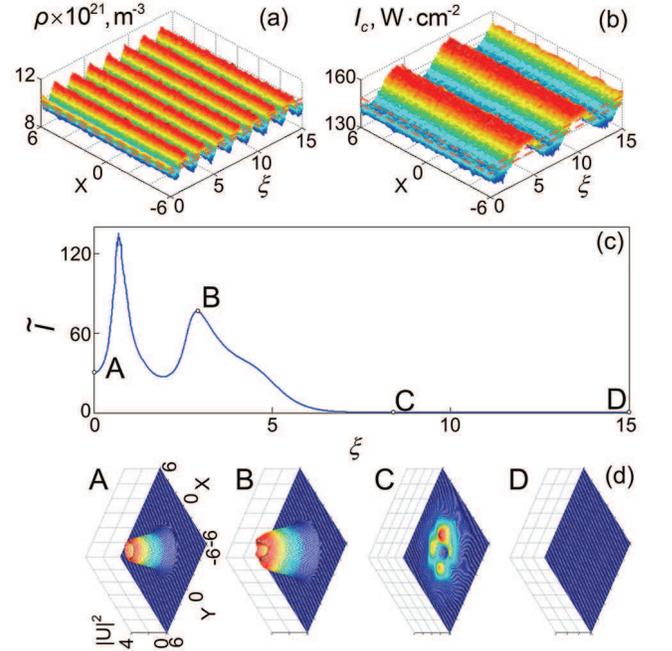}
\caption{\label{fig:10}Model of the loss of stability as result
periodic modulation of control parameters (a), (b) in the presence of
a low-intensity noise. Image (c) shows the temporal dynamics
of $\tilde{I}$; the probe beam shapes corresponding to
different timepoints from (c) are presented in (d).}
\end{figure}

To study the effect of perturbations on the vortex dynamics we
performed a series of numerical experiments for different values of
$\sigma_I$ and $\sigma_\rho$ and the control parameters $I_{c}^{\rm
det}$ and $\rho^{\rm det}$. As an indirect criterium of stability we
assume the statistical stability of the reduced power
\[\tilde{I}={\rm \;}\int_{0}^{2\pi }\int_{0}^{\infty }\left|U(r,\theta )\right|^{2} rdrd\theta\]
shown in Figs.~\ref{fig:8}c-\ref{fig:10}c under perturbations
$\zeta_{I}(X,Y,\xi)$ and $\zeta_{\rho}(X,Y,\xi)$. If stabilization
of $\tilde{I}$ is observed, then we restore and analyze numerically
the shape of the soliton in the traced point on axis $\xi$.

The longitudinal perturbations give rise to the inert properties of
the soliton, i.e., the presence of noise, in which the phase
trajectories $\rho(X,Y,\xi),I_{c}(X,Y,\xi)$ never leave the
stability region I in Fig.~\ref{fig:5}, does not cause dramatic
consequences. Beyond the stability area the dynamics of a vortex
is determined by the retardation effects in slow transformations its shape
compared to the rapid changes in GLE parameters initiated by the
noise. Figure~\ref{fig:8} shows the case of the critical
noise parameters $\varepsilon_{I}^{cr}=5.1\%$ and
$\varepsilon_{\rho}^{cr}=7.4\%$ ($I_{c}^{\rm det}={\rm
146.5}$~W$\cdot$cm$^{-2}$ and $\rho^{\rm det}=10.1\times 10^{21}$ m$^{-3}$
correspond to point A in Fig.~\ref{fig:5}) for the vortex soliton
which preserves its shape over long distances $z=16.1$~m at
$\xi=10000$  for values of the
noise strength $D_{\mu}=3.2\times 10^{-3}$, $D_{\nu} =1.1\times
10^{-3}$, $D_{\delta}=1.66\times 10^{-4}$, $D_{\phi} =1.1\times
10^{-3} $. The increase in parameters $\varepsilon_{I}^{cr}$ and $\varepsilon_{\rho}^{cr}$ on $0.1\%$ leads to fast damping of a soliton.
The presence of the transverse perturbations in plane $xOy$ leads to the noise-induced distortion of the wave packet
profile in Fig.~\ref{fig:8}d but has almost no effect
on its stability.

The noise contribution in parameters of Eq.~(\ref{eq:24}) for a regime with
$I_c^{\rm det}=146$~W$\cdot$cm$^{-2}$ and $\rho^{\rm det}=9.77\times
10^{21}$~m$^{-3}$ from the transitional area II in Fig.~\ref{fig:5}
can initiate reduction of topology and a jump transition of vortex
solitons in a ``breathing'' mode in the time point $\xi\cong1000$ in Fig.~\ref{fig:9}c. This regime is characterized by constant interconversion
between the fundamental soliton and oscillating polygonal structure
as on images C, D in Fig.~\ref{fig:9}d. The observed evolution is the result of migration the
optical structure performs between areas III and V in
Fig.~\ref{fig:5} for values of noise strength $D_{\mu}=1.8\times
10^{-3}$, $D_{\nu}=7.23\times 10^{-4}$, $D_{\delta} =1.2\times
10^{-4}$, $D_{\phi}=7.4\times 10^{-4}$ which correspond to the
relative deviations $\varepsilon_I=1.92\%$ and $\varepsilon_{\rho}=2.87\%$ at $\xi_{\rm rnd}=0.05$. In the process of transition of a soliton to
the vortex-free mode one may observe formation of crescent-shaped
optical structure as shown on image B in Fig.~\ref{fig:9}d. In further rising
of $\varepsilon_I$ and $\varepsilon_{\rho}$ there comes a transition into an
unstable mode and decay, which is associated with a disproportionate
increase in the noise strength for various GLE coefficients. In
this model for $\Lambda$-scheme of interaction the fifth-order nonlinear absorption coefficient $\mu$
has a more rapid increase in the noise strength and brings the soliton
to zone IV in Fig.~\ref{fig:5} and destroys it there.

The destructive effect are also caused by a weak periodic modulation
of control parameters in the presence of a low intensity noise in
Fig.~\ref{fig:10}a,b, which collectively lead to fast violations of
dynamic equilibrium in the system and damping of the soliton at
scales of the order of $\xi =8$ in Fig.~\ref{fig:10}c,d. The simulation parameters
correspond to $\zeta_{I_0}^{\rm reg}=7.5$~W$\cdot$cm$^{-2}$, $\varphi_{I}
=-2.51$ and $\zeta_{\rho_0}^{\rm reg}=0.75\times 10^{21}$~m$^{-3}$,
$\varphi_{\rho}=2.76$, $L_I=6$, $L_\rho=2$ at average values of
$I_{c}^{\rm det}$ and $\rho^{\rm det}$ as for point A in
Fig.~\ref{fig:5}. The scale of density fluctuations in
Fig.~\ref{fig:10}a is $3.2$~mm that given the expression for the
speed of sound in a Bose gas $c_{sn}=(\hbar/m_{0})\sqrt{4\pi
a_a\rho}$ \cite{Stat2} corresponds to $c_{sn}=3.2$~mm/s for the sound wave with the frequency $1$~Hz. Here $m_0$ is the mass of a
single atom and $a_{a}$ is the scattering amplitude. In the process of damping of the vortex soliton the system shows the rapid development of angular perturbations leading
to the emergence of two or multihump unstable localized structures as for images B, C in Fig.~\ref{fig:10}d.
This behavior is associated with a large spatial scale of the
simulated periodic perturbations which keep the vortex in unstable area for a long time.

The well-known estimates for the modern laser systems that are
compatible with the telecommunication channels determine the maximum
allowable noise level of the output intensity to be not higher than
$1\%$. This estimation by a large margin falls within the considered theory of
stability of vortex solitons (compare with $\varepsilon_{I}^{cr}$ in Fig.~\ref{fig:8}). Particle number fluctuations of a Bose
gas in the range of ${\rm T}<{\rm T}_{{\rm cr}}$ have the
temperature dependence \cite{Politzer}
\[\sigma_{N}^{\rm 2}=\frac{\pi^{2}}{6 \zeta(3)}N\left(\frac{T}{T_{cr}}\right)^{3}.\]
This equation provides a simple expression for estimating the maximum-allowable density fluctuation in BEC $\varepsilon_{\rho}^{max}=
\pi\left(6\zeta(3)N\right)^{-1/2}\cong1.17 N^{-1/2}$ under condition that $T=T_{cr}$. For $N=3\times10^{4}$ and $\omega_{0}=50$ kHz the critical temperature for gas of $^{87}\rm Rb$ in the fiber is $72$ $\mu$K. The maximum-allowable relative density fluctuations in such a system corresponds to $\varepsilon_{\rho}^{max}=0.68\%$, which is $10$ times as low as the critical atomic noise calculated for vortex soliton in Fig.~\ref{fig:8}. This fact makes it evident to observe the vortex soliton regime at a temperature below the critical temperature for BEC.

In numerical calculations for Figs.~\ref{fig:8}-\ref{fig:10} we also consider the fluctuations
of diffraction parameter emerging in the cross-section with the
spatial scale within the range of the field's wavelength. This
requirement arises from the fact that the present fluctuations of
the gas density effectively modulate the refractive index of the
medium. They also create a random phase delay, which with the accounting for the finite width
of the emission line produce the wavelength noise parameter. However, the simulation shows a statistically stable picture
with the presence of local perturbations of soliton shapes even at
${2\%}$ deviations of $\lambda$. This is similar to the initial
angular perturbations of the vortex shapes which are smoothed after
elimination of the noise.

\section{Conclusion}
In this paper we have studied the influence of optical and temperature-dependent atomic fluctuations on the formation and propagation of optical vortex solitons in dense media realized as hollow-core optical fibers filled with a cold atomic gas in presence of optical pumping. This investigation may give some basic provisions that could be useful for development of optical communication channels using vortex solitons in hollow-core fibers filled with a cold atomic gas.  With the given parameters of the fiber and the proper intensity of the probe field one can simultaneously tune the intensity and frequency of the cw pump field to provide the conditions for the soliton propagation regime of vortex structures in the optical system. Because of the nonlinear contribution from the local field effects to development of the competitive optical processes in a dense atomic medium it is necessary to account for the NDD corrections in calculating the values of the control parameters for which we expect appearance of the solitons. For the parameters used in the direct numerical simulations the maximum fluctuations of the pump intensity and the probe wavelength for which the soliton regime is still valid impose minor restrictions on the stability of the pump laser power and the probe laser frequency. The estimates for the atomic fluctuations gives an opportunity to observe the optical  vortex solitions in the core-filling gas of the fiber for temperatures smaller than the critical temperature for BEC. The large-scale perturbations in the development of acoustic waves and the modulations of the pump field in the fiber can quickly destroy the soliton propagation regime even if they are of small amplitude.

In practice, for long-term maintenance of the BEC state in the information
channel it is possible to use fine fibers filled with a cold atoms
\cite{Balykin:1996}. Longer lifetimes of a coherent state in such a
system are due to the geometry of the fiber and, in addition, the
effect of channeling of atoms at creating the surface light wave
along its core, causing the atoms to lose their energy effectively.
The estimates given in \cite{Balykin:1996} show that for the density
$10^{15}$~cm$^{-3}$ the temperature inside the fiber drops
down to $1.5\times 10^{-5}$~K.

However, such an elongated BEC must be assumed as a $1$D gas so it is necessary to use some different relations for the fluctuations \cite{Petrov} at which observation of the vortex solitons is possible. Except it, the use of thin hollow-core fibers for optical
communication due to the need to create localized optical structures
at nanoscale, but in solving such a problem transition to the
Maxwell equations for continuous media is impossible, and the
effects of the near-field will play a crucial role.

\section{ACKNOWLEDGMENTS}
This work was supported by Russian Foundation for Basic Research Grant No. 12-02-31573 mol\underline{ }a and the Ministry of Education and Science of the Russian Federation under the program ``Scientific and scientific-pedagogical potential of Russia for innovation" (Project No. 14.132.21.1397).

\bibliography{Prokhorov}

\begin{thebibliography}{46}%
\makeatletter
\providecommand \@ifxundefined [1]{%
 \@ifx{#1\undefined}
}%
\providecommand \@ifnum [1]{%
 \ifnum #1\expandafter \@firstoftwo
 \else \expandafter \@secondoftwo
 \fi
}%
\providecommand \@ifx [1]{%
 \ifx #1\expandafter \@firstoftwo
 \else \expandafter \@secondoftwo
 \fi
}%
\providecommand \natexlab [1]{#1}%
\providecommand \enquote  [1]{``#1''}%
\providecommand \bibnamefont  [1]{#1}%
\providecommand \bibfnamefont [1]{#1}%
\providecommand \citenamefont [1]{#1}%
\providecommand \href@noop [0]{\@secondoftwo}%
\providecommand \href [0]{\begingroup \@sanitize@url \@href}%
\providecommand \@href[1]{\@@startlink{#1}\@@href}%
\providecommand \@@href[1]{\endgroup#1\@@endlink}%
\providecommand \@sanitize@url [0]{\catcode `\\12\catcode `\$12\catcode
  `\&12\catcode `\#12\catcode `\^12\catcode `\_12\catcode `\%12\relax}%
\providecommand \@@startlink[1]{}%
\providecommand \@@endlink[0]{}%
\providecommand \url  [0]{\begingroup\@sanitize@url \@url }%
\providecommand \@url [1]{\endgroup\@href {#1}{\urlprefix }}%
\providecommand \urlprefix  [0]{URL }%
\providecommand \Eprint [0]{\href }%
\providecommand \doibase [0]{http://dx.doi.org/}%
\providecommand \selectlanguage [0]{\@gobble}%
\providecommand \bibinfo  [0]{\@secondoftwo}%
\providecommand \bibfield  [0]{\@secondoftwo}%
\providecommand \translation [1]{[#1]}%
\providecommand \BibitemOpen [0]{}%
\providecommand \bibitemStop [0]{}%
\providecommand \bibitemNoStop [0]{.\EOS\space}%
\providecommand \EOS [0]{\spacefactor3000\relax}%
\providecommand \BibitemShut  [1]{\csname bibitem#1\endcsname}%
\let\auto@bib@innerbib\@empty
\bibitem [{\citenamefont {Kivshar}\ and\ \citenamefont
  {Agrawal}(2003)}]{Kivshar}%
  \BibitemOpen
  \bibfield  {author} {\bibinfo {author} {\bibfnamefont {Y.~S.}\ \bibnamefont
  {Kivshar}}\ and\ \bibinfo {author} {\bibfnamefont {G.}~\bibnamefont
  {Agrawal}},\ }\href@noop {} {\emph {\bibinfo {title} {Optical Solitons: From
  Fibers to Photonic Crystals}}}\ (\bibinfo  {publisher} {Academic Press},\
  \bibinfo {year} {2003})\BibitemShut {NoStop}%
\bibitem [{\citenamefont {Rosanov}(2011)}]{Rozanov}%
  \BibitemOpen
  \bibfield  {author} {\bibinfo {author} {\bibfnamefont {N.~N.}\ \bibnamefont
  {Rosanov}},\ }\href@noop {} {\emph {\bibinfo {title} {Dissipative optical
  solitons: from Micro- to Nano- and Attosolitons}}}\ (\bibinfo  {publisher}
  {FIZMATLIT},\ \bibinfo {year} {2011})\BibitemShut {NoStop}%
\bibitem [{\citenamefont {Skarka}\ \emph
  {et~al.}(2010{\natexlab{a}})\citenamefont {Skarka}, \citenamefont
  {Aleksi\ifmmode~\acute{c}\else \'{c}\fi{}}, \citenamefont {Leblond},
  \citenamefont {Malomed},\ and\ \citenamefont
  {Mihalache}}]{PhysRevLett.105.213901}%
  \BibitemOpen
  \bibfield  {author} {\bibinfo {author} {\bibfnamefont {V.}~\bibnamefont
  {Skarka}}, \bibinfo {author} {\bibfnamefont {N.~B.}\ \bibnamefont
  {Aleksi\ifmmode~\acute{c}\else \'{c}\fi{}}}, \bibinfo {author} {\bibfnamefont
  {H.}~\bibnamefont {Leblond}}, \bibinfo {author} {\bibfnamefont {B.~A.}\
  \bibnamefont {Malomed}}, \ and\ \bibinfo {author} {\bibfnamefont
  {D.}~\bibnamefont {Mihalache}},\ }\href {\doibase
  10.1103/PhysRevLett.105.213901} {\bibfield  {journal} {\bibinfo  {journal}
  {Phys. Rev. Lett.}\ }\textbf {\bibinfo {volume} {105}},\ \bibinfo {pages}
  {213901} (\bibinfo {year} {2010}{\natexlab{a}})}\BibitemShut {NoStop}%
\bibitem [{\citenamefont {Pugatch}\ \emph {et~al.}(2007)\citenamefont
  {Pugatch}, \citenamefont {Shuker}, \citenamefont {Firstenberg}, \citenamefont
  {Ron},\ and\ \citenamefont {Davidson}}]{PhysRevLett.98.203601}%
  \BibitemOpen
  \bibfield  {author} {\bibinfo {author} {\bibfnamefont {R.}~\bibnamefont
  {Pugatch}}, \bibinfo {author} {\bibfnamefont {M.}~\bibnamefont {Shuker}},
  \bibinfo {author} {\bibfnamefont {O.}~\bibnamefont {Firstenberg}}, \bibinfo
  {author} {\bibfnamefont {A.}~\bibnamefont {Ron}}, \ and\ \bibinfo {author}
  {\bibfnamefont {N.}~\bibnamefont {Davidson}},\ }\href {\doibase
  10.1103/PhysRevLett.98.203601} {\bibfield  {journal} {\bibinfo  {journal}
  {Phys. Rev. Lett.}\ }\textbf {\bibinfo {volume} {98}},\ \bibinfo {pages}
  {203601} (\bibinfo {year} {2007})}\BibitemShut {NoStop}%
\bibitem [{\citenamefont {Smith}\ and\ \citenamefont
  {Armstrong}(2003)}]{Smith:03}%
  \BibitemOpen
  \bibfield  {author} {\bibinfo {author} {\bibfnamefont {A.}~\bibnamefont
  {Smith}}\ and\ \bibinfo {author} {\bibfnamefont {D.}~\bibnamefont
  {Armstrong}},\ }\href {\doibase 10.1364/OE.11.000868} {\bibfield  {journal}
  {\bibinfo  {journal} {Opt. Express}\ }\textbf {\bibinfo {volume} {11}},\
  \bibinfo {pages} {868} (\bibinfo {year} {2003})}\BibitemShut {NoStop}%
\bibitem [{\citenamefont {Bezuhanov}\ \emph {et~al.}(2004)\citenamefont
  {Bezuhanov}, \citenamefont {Dreischuh}, \citenamefont {Paulus}, \citenamefont
  {Sch\"{a}tzel},\ and\ \citenamefont {Walther}}]{Bezuhanov:04}%
  \BibitemOpen
  \bibfield  {author} {\bibinfo {author} {\bibfnamefont {K.}~\bibnamefont
  {Bezuhanov}}, \bibinfo {author} {\bibfnamefont {A.}~\bibnamefont
  {Dreischuh}}, \bibinfo {author} {\bibfnamefont {G.~G.}\ \bibnamefont
  {Paulus}}, \bibinfo {author} {\bibfnamefont {M.~G.}\ \bibnamefont
  {Sch\"{a}tzel}}, \ and\ \bibinfo {author} {\bibfnamefont {H.}~\bibnamefont
  {Walther}},\ }\href {\doibase 10.1364/OL.29.001942} {\bibfield  {journal}
  {\bibinfo  {journal} {Opt. Lett.}\ }\textbf {\bibinfo {volume} {29}},\
  \bibinfo {pages} {1942} (\bibinfo {year} {2004})}\BibitemShut {NoStop}%
\bibitem [{\citenamefont {Genevet}\ \emph {et~al.}(2010)\citenamefont
  {Genevet}, \citenamefont {Barland}, \citenamefont {Giudici},\ and\
  \citenamefont {Tredicce}}]{PhysRevLett.104.223902}%
  \BibitemOpen
  \bibfield  {author} {\bibinfo {author} {\bibfnamefont {P.}~\bibnamefont
  {Genevet}}, \bibinfo {author} {\bibfnamefont {S.}~\bibnamefont {Barland}},
  \bibinfo {author} {\bibfnamefont {M.}~\bibnamefont {Giudici}}, \ and\
  \bibinfo {author} {\bibfnamefont {J.~R.}\ \bibnamefont {Tredicce}},\ }\href
  {\doibase 10.1103/PhysRevLett.104.223902} {\bibfield  {journal} {\bibinfo
  {journal} {Phys. Rev. Lett.}\ }\textbf {\bibinfo {volume} {104}},\ \bibinfo
  {pages} {223902} (\bibinfo {year} {2010})}\BibitemShut {NoStop}%
\bibitem [{\citenamefont {Prokhorov}\ \emph {et~al.}(2012)\citenamefont
  {Prokhorov}, \citenamefont {Gubin}, \citenamefont {Leksin}, \citenamefont
  {Gladush}, \citenamefont {Alodjants},\ and\ \citenamefont
  {Arakelian}}]{JEPT:2012}%
  \BibitemOpen
  \bibfield  {author} {\bibinfo {author} {\bibfnamefont {A.}~\bibnamefont
  {Prokhorov}}, \bibinfo {author} {\bibfnamefont {M.}~\bibnamefont {Gubin}},
  \bibinfo {author} {\bibfnamefont {A.}~\bibnamefont {Leksin}}, \bibinfo
  {author} {\bibfnamefont {M.}~\bibnamefont {Gladush}}, \bibinfo {author}
  {\bibfnamefont {A.}~\bibnamefont {Alodjants}}, \ and\ \bibinfo {author}
  {\bibfnamefont {S.}~\bibnamefont {Arakelian}},\ }\href {\doibase
  10.1134/S1063776112050111} {\bibfield  {journal} {\bibinfo  {journal}
  {Journal of Experimental and Theoretical Physics}\ }\textbf {\bibinfo
  {volume} {115}},\ \bibinfo {pages} {1} (\bibinfo {year} {2012})}\BibitemShut
  {NoStop}%
\bibitem [{\citenamefont {Gorbach}\ \emph {et~al.}(2008)\citenamefont
  {Gorbach}, \citenamefont {Skryabin},\ and\ \citenamefont
  {Harvey}}]{PhysRevA.77.063810}%
  \BibitemOpen
  \bibfield  {author} {\bibinfo {author} {\bibfnamefont {A.~V.}\ \bibnamefont
  {Gorbach}}, \bibinfo {author} {\bibfnamefont {D.~V.}\ \bibnamefont
  {Skryabin}}, \ and\ \bibinfo {author} {\bibfnamefont {C.~N.}\ \bibnamefont
  {Harvey}},\ }\href {\doibase 10.1103/PhysRevA.77.063810} {\bibfield
  {journal} {\bibinfo  {journal} {Phys. Rev. A}\ }\textbf {\bibinfo {volume}
  {77}},\ \bibinfo {pages} {063810} (\bibinfo {year} {2008})}\BibitemShut
  {NoStop}%
\bibitem [{\citenamefont {Herman}\ \emph {et~al.}(1989)\citenamefont {Herman},
  \citenamefont {Eberly},\ and\ \citenamefont {Raymer}}]{PhysRevA.39.3447}%
  \BibitemOpen
  \bibfield  {author} {\bibinfo {author} {\bibfnamefont {B.~J.}\ \bibnamefont
  {Herman}}, \bibinfo {author} {\bibfnamefont {J.~H.}\ \bibnamefont {Eberly}},
  \ and\ \bibinfo {author} {\bibfnamefont {M.~G.}\ \bibnamefont {Raymer}},\
  }\href {\doibase 10.1103/PhysRevA.39.3447} {\bibfield  {journal} {\bibinfo
  {journal} {Phys. Rev. A}\ }\textbf {\bibinfo {volume} {39}},\ \bibinfo
  {pages} {3447} (\bibinfo {year} {1989})}\BibitemShut {NoStop}%
\bibitem [{\citenamefont {Gorshkov}\ \emph {et~al.}(2007)\citenamefont
  {Gorshkov}, \citenamefont {Andr\'e}, \citenamefont {Lukin},\ and\
  \citenamefont {S\o{}rensen}}]{PhysRevA.76.033805}%
  \BibitemOpen
  \bibfield  {author} {\bibinfo {author} {\bibfnamefont {A.~V.}\ \bibnamefont
  {Gorshkov}}, \bibinfo {author} {\bibfnamefont {A.}~\bibnamefont {Andr\'e}},
  \bibinfo {author} {\bibfnamefont {M.~D.}\ \bibnamefont {Lukin}}, \ and\
  \bibinfo {author} {\bibfnamefont {A.~S.}\ \bibnamefont {S\o{}rensen}},\
  }\href {\doibase 10.1103/PhysRevA.76.033805} {\bibfield  {journal} {\bibinfo
  {journal} {Phys. Rev. A}\ }\textbf {\bibinfo {volume} {76}},\ \bibinfo
  {pages} {033805} (\bibinfo {year} {2007})}\BibitemShut {NoStop}%
\bibitem [{\citenamefont {Bajcsy}\ \emph {et~al.}(2011)\citenamefont {Bajcsy},
  \citenamefont {Hofferberth}, \citenamefont {Peyronel}, \citenamefont {Balic},
  \citenamefont {Liang}, \citenamefont {Zibrov}, \citenamefont {Vuletic},\ and\
  \citenamefont {Lukin}}]{PhysRevA.83.063830}%
  \BibitemOpen
  \bibfield  {author} {\bibinfo {author} {\bibfnamefont {M.}~\bibnamefont
  {Bajcsy}}, \bibinfo {author} {\bibfnamefont {S.}~\bibnamefont {Hofferberth}},
  \bibinfo {author} {\bibfnamefont {T.}~\bibnamefont {Peyronel}}, \bibinfo
  {author} {\bibfnamefont {V.}~\bibnamefont {Balic}}, \bibinfo {author}
  {\bibfnamefont {Q.}~\bibnamefont {Liang}}, \bibinfo {author} {\bibfnamefont
  {A.~S.}\ \bibnamefont {Zibrov}}, \bibinfo {author} {\bibfnamefont
  {V.}~\bibnamefont {Vuletic}}, \ and\ \bibinfo {author} {\bibfnamefont
  {M.~D.}\ \bibnamefont {Lukin}},\ }\href {\doibase 10.1103/PhysRevA.83.063830}
  {\bibfield  {journal} {\bibinfo  {journal} {Phys. Rev. A}\ }\textbf {\bibinfo
  {volume} {83}},\ \bibinfo {pages} {063830} (\bibinfo {year}
  {2011})}\BibitemShut {NoStop}%
\bibitem [{\citenamefont {Fedotov}\ \emph {et~al.}(2004)\citenamefont
  {Fedotov}, \citenamefont {Konorov}, \citenamefont {Mitrokhin}, \citenamefont
  {Serebryannikov},\ and\ \citenamefont {Zheltikov}}]{Zheltikov}%
  \BibitemOpen
  \bibfield  {author} {\bibinfo {author} {\bibfnamefont {A.~B.}\ \bibnamefont
  {Fedotov}}, \bibinfo {author} {\bibfnamefont {S.~O.}\ \bibnamefont
  {Konorov}}, \bibinfo {author} {\bibfnamefont {V.~P.}\ \bibnamefont
  {Mitrokhin}}, \bibinfo {author} {\bibfnamefont {E.~E.}\ \bibnamefont
  {Serebryannikov}}, \ and\ \bibinfo {author} {\bibfnamefont {A.~M.}\
  \bibnamefont {Zheltikov}},\ }\href@noop {} {\bibfield  {journal} {\bibinfo
  {journal} {Phys. Rev. A}\ }\textbf {\bibinfo {volume} {70}},\ \bibinfo
  {pages} {045802} (\bibinfo {year} {2004})}\BibitemShut {NoStop}%
\bibitem [{\citenamefont {Serebryannikov}\ \emph {et~al.}(2006)\citenamefont
  {Serebryannikov}, \citenamefont {Zheltikov}, \citenamefont {K\"{o}hler},
  \citenamefont {Ishii}, \citenamefont {Teisset}, \citenamefont {Fuji},
  \citenamefont {Krausz},\ and\ \citenamefont {Baltu\u{s}ka}}]{Kohler}%
  \BibitemOpen
  \bibfield  {author} {\bibinfo {author} {\bibfnamefont {E.~E.}\ \bibnamefont
  {Serebryannikov}}, \bibinfo {author} {\bibfnamefont {A.~M.}\ \bibnamefont
  {Zheltikov}}, \bibinfo {author} {\bibfnamefont {S.}~\bibnamefont
  {K\"{o}hler}}, \bibinfo {author} {\bibfnamefont {N.}~\bibnamefont {Ishii}},
  \bibinfo {author} {\bibfnamefont {C.~Y.}\ \bibnamefont {Teisset}}, \bibinfo
  {author} {\bibfnamefont {T.}~\bibnamefont {Fuji}}, \bibinfo {author}
  {\bibfnamefont {F.}~\bibnamefont {Krausz}}, \ and\ \bibinfo {author}
  {\bibfnamefont {A.}~\bibnamefont {Baltu\u{s}ka}},\ }\href@noop {} {\bibfield
  {journal} {\bibinfo  {journal} {Phys. Rev. E}\ }\textbf {\bibinfo {volume}
  {73}},\ \bibinfo {pages} {066617} (\bibinfo {year} {2006})}\BibitemShut
  {NoStop}%
\bibitem [{\citenamefont {Essiambre}\ \emph {et~al.}(2008)\citenamefont
  {Essiambre}, \citenamefont {Foschini}, \citenamefont {Kramer},\ and\
  \citenamefont {Winzer}}]{PhysRevLett.101.163901}%
  \BibitemOpen
  \bibfield  {author} {\bibinfo {author} {\bibfnamefont {R.-J.}\ \bibnamefont
  {Essiambre}}, \bibinfo {author} {\bibfnamefont {G.~J.}\ \bibnamefont
  {Foschini}}, \bibinfo {author} {\bibfnamefont {G.}~\bibnamefont {Kramer}}, \
  and\ \bibinfo {author} {\bibfnamefont {P.~J.}\ \bibnamefont {Winzer}},\
  }\href {\doibase 10.1103/PhysRevLett.101.163901} {\bibfield  {journal}
  {\bibinfo  {journal} {Phys. Rev. Lett.}\ }\textbf {\bibinfo {volume} {101}},\
  \bibinfo {pages} {163901} (\bibinfo {year} {2008})}\BibitemShut {NoStop}%
\bibitem [{\citenamefont {Hopf}\ \emph {et~al.}(1984)\citenamefont {Hopf},
  \citenamefont {Bowden},\ and\ \citenamefont {Louisell}}]{PhysRevA.29.2591}%
  \BibitemOpen
  \bibfield  {author} {\bibinfo {author} {\bibfnamefont {F.~A.}\ \bibnamefont
  {Hopf}}, \bibinfo {author} {\bibfnamefont {C.~M.}\ \bibnamefont {Bowden}}, \
  and\ \bibinfo {author} {\bibfnamefont {W.~H.}\ \bibnamefont {Louisell}},\
  }\href {\doibase 10.1103/PhysRevA.29.2591} {\bibfield  {journal} {\bibinfo
  {journal} {Phys. Rev. A}\ }\textbf {\bibinfo {volume} {29}},\ \bibinfo
  {pages} {2591} (\bibinfo {year} {1984})}\BibitemShut {NoStop}%
\bibitem [{\citenamefont {Kuznetsov}\ \emph {et~al.}(2011)\citenamefont
  {Kuznetsov}, \citenamefont {Roerich},\ and\ \citenamefont
  {Gladush}}]{GladJETP:2011}%
  \BibitemOpen
  \bibfield  {author} {\bibinfo {author} {\bibfnamefont {D.}~\bibnamefont
  {Kuznetsov}}, \bibinfo {author} {\bibfnamefont {V.}~\bibnamefont {Roerich}},
  \ and\ \bibinfo {author} {\bibfnamefont {M.}~\bibnamefont {Gladush}},\ }\href
  {\doibase 10.1134/S1063776111100050} {\bibfield  {journal} {\bibinfo
  {journal} {Journal of Experimental and Theoretical Physics}\ }\textbf
  {\bibinfo {volume} {113}},\ \bibinfo {pages} {647} (\bibinfo {year}
  {2011})}\BibitemShut {NoStop}%
\bibitem [{\citenamefont {Vlasov}\ and\ \citenamefont
  {Lemeza}(2011)}]{PhysRevA.84.023828}%
  \BibitemOpen
  \bibfield  {author} {\bibinfo {author} {\bibfnamefont {R.~A.}\ \bibnamefont
  {Vlasov}}\ and\ \bibinfo {author} {\bibfnamefont {A.~M.}\ \bibnamefont
  {Lemeza}},\ }\href {\doibase 10.1103/PhysRevA.84.023828} {\bibfield
  {journal} {\bibinfo  {journal} {Phys. Rev. A}\ }\textbf {\bibinfo {volume}
  {84}},\ \bibinfo {pages} {023828} (\bibinfo {year} {2011})}\BibitemShut
  {NoStop}%
\bibitem [{\citenamefont {Crenshaw}(2008)}]{PhysRevA.78.053827}%
  \BibitemOpen
  \bibfield  {author} {\bibinfo {author} {\bibfnamefont {M.~E.}\ \bibnamefont
  {Crenshaw}},\ }\href {\doibase 10.1103/PhysRevA.78.053827} {\bibfield
  {journal} {\bibinfo  {journal} {Phys. Rev. A}\ }\textbf {\bibinfo {volume}
  {78}},\ \bibinfo {pages} {053827} (\bibinfo {year} {2008})}\BibitemShut
  {NoStop}%
\bibitem [{\citenamefont {Dolgaleva}\ and\ \citenamefont
  {Boyd}(2012)}]{Dolgaleva:12}%
  \BibitemOpen
  \bibfield  {author} {\bibinfo {author} {\bibfnamefont {K.}~\bibnamefont
  {Dolgaleva}}\ and\ \bibinfo {author} {\bibfnamefont {R.~W.}\ \bibnamefont
  {Boyd}},\ }\href {\doibase 10.1364/AOP.4.000001} {\bibfield  {journal}
  {\bibinfo  {journal} {Adv. Opt. Photon.}\ }\textbf {\bibinfo {volume} {4}},\
  \bibinfo {pages} {1} (\bibinfo {year} {2012})}\BibitemShut {NoStop}%
\bibitem [{\citenamefont {Mollenauer}\ \emph {et~al.}(1996)\citenamefont
  {Mollenauer}, \citenamefont {Mamyshev},\ and\ \citenamefont
  {Neubelt}}]{Mollenauer:96}%
  \BibitemOpen
  \bibfield  {author} {\bibinfo {author} {\bibfnamefont {L.~F.}\ \bibnamefont
  {Mollenauer}}, \bibinfo {author} {\bibfnamefont {P.~V.}\ \bibnamefont
  {Mamyshev}}, \ and\ \bibinfo {author} {\bibfnamefont {M.~J.}\ \bibnamefont
  {Neubelt}},\ }\href {\doibase 10.1364/OL.21.001724} {\bibfield  {journal}
  {\bibinfo  {journal} {Opt. Lett.}\ }\textbf {\bibinfo {volume} {21}},\
  \bibinfo {pages} {1724} (\bibinfo {year} {1996})}\BibitemShut {NoStop}%
\bibitem [{\citenamefont {Chertkov}\ \emph {et~al.}(2002)\citenamefont
  {Chertkov}, \citenamefont {Gabitov}, \citenamefont {Lushnikov}, \citenamefont
  {Moeser},\ and\ \citenamefont {Toroczkai}}]{Chertkov:02}%
  \BibitemOpen
  \bibfield  {author} {\bibinfo {author} {\bibfnamefont {M.}~\bibnamefont
  {Chertkov}}, \bibinfo {author} {\bibfnamefont {I.}~\bibnamefont {Gabitov}},
  \bibinfo {author} {\bibfnamefont {P.~M.}\ \bibnamefont {Lushnikov}}, \bibinfo
  {author} {\bibfnamefont {J.}~\bibnamefont {Moeser}}, \ and\ \bibinfo {author}
  {\bibfnamefont {Z.}~\bibnamefont {Toroczkai}},\ }\href {\doibase
  10.1364/JOSAB.19.002538} {\bibfield  {journal} {\bibinfo  {journal} {J. Opt.
  Soc. Am. B}\ }\textbf {\bibinfo {volume} {19}},\ \bibinfo {pages} {2538}
  (\bibinfo {year} {2002})}\BibitemShut {NoStop}%
\bibitem [{\citenamefont {Graham}\ and\ \citenamefont
  {T\'el}(1990)}]{PhysRevA.42.4661}%
  \BibitemOpen
  \bibfield  {author} {\bibinfo {author} {\bibfnamefont {R.}~\bibnamefont
  {Graham}}\ and\ \bibinfo {author} {\bibfnamefont {T.}~\bibnamefont {T\'el}},\
  }\href {\doibase 10.1103/PhysRevA.42.4661} {\bibfield  {journal} {\bibinfo
  {journal} {Phys. Rev. A}\ }\textbf {\bibinfo {volume} {42}},\ \bibinfo
  {pages} {4661} (\bibinfo {year} {1990})}\BibitemShut {NoStop}%
\bibitem [{\citenamefont {Cartes}\ \emph {et~al.}(2012)\citenamefont {Cartes},
  \citenamefont {Descalzi},\ and\ \citenamefont {Brand}}]{PhysRevE.85.015205}%
  \BibitemOpen
  \bibfield  {author} {\bibinfo {author} {\bibfnamefont {C.}~\bibnamefont
  {Cartes}}, \bibinfo {author} {\bibfnamefont {O.}~\bibnamefont {Descalzi}}, \
  and\ \bibinfo {author} {\bibfnamefont {H.~R.}\ \bibnamefont {Brand}},\ }\href
  {\doibase 10.1103/PhysRevE.85.015205} {\bibfield  {journal} {\bibinfo
  {journal} {Phys. Rev. E}\ }\textbf {\bibinfo {volume} {85}},\ \bibinfo
  {pages} {015205} (\bibinfo {year} {2012})}\BibitemShut {NoStop}%
\bibitem [{\citenamefont {Einstein}(1905)}]{Einstein:1905}%
  \BibitemOpen
  \bibfield  {author} {\bibinfo {author} {\bibfnamefont {A.}~\bibnamefont
  {Einstein}},\ }\href {\doibase 10.1002/andp.19053220806} {\bibfield
  {journal} {\bibinfo  {journal} {Annalen der Physik}\ }\textbf {\bibinfo
  {volume} {322}},\ \bibinfo {pages} {549} (\bibinfo {year}
  {1905})}\BibitemShut {NoStop}%
\bibitem [{\citenamefont {von Smoluchowski}(1906)}]{Smoluchowski:1906}%
  \BibitemOpen
  \bibfield  {author} {\bibinfo {author} {\bibfnamefont {M.}~\bibnamefont {von
  Smoluchowski}},\ }\href {\doibase 10.1002/andp.19063261405} {\bibfield
  {journal} {\bibinfo  {journal} {Annalen der Physik}\ }\textbf {\bibinfo
  {volume} {326}},\ \bibinfo {pages} {756} (\bibinfo {year}
  {1906})}\BibitemShut {NoStop}%
\bibitem [{\citenamefont {Giorgini}\ \emph {et~al.}(1998)\citenamefont
  {Giorgini}, \citenamefont {Pitaevskii},\ and\ \citenamefont
  {Stringari}}]{Giorgini}%
  \BibitemOpen
  \bibfield  {author} {\bibinfo {author} {\bibfnamefont {S.}~\bibnamefont
  {Giorgini}}, \bibinfo {author} {\bibfnamefont {L.~P.}\ \bibnamefont
  {Pitaevskii}}, \ and\ \bibinfo {author} {\bibfnamefont {S.}~\bibnamefont
  {Stringari}},\ }\href@noop {} {\bibfield  {journal} {\bibinfo  {journal}
  {Phys. Rev. Lett.}\ }\textbf {\bibinfo {volume} {80}},\ \bibinfo {pages}
  {5040} (\bibinfo {year} {1998})}\BibitemShut {NoStop}%
\bibitem [{\citenamefont {Kocharovsky}\ \emph {et~al.}(2000)\citenamefont
  {Kocharovsky}, \citenamefont {Kocharovsky},\ and\ \citenamefont
  {Scully}}]{Kocharovsky}%
  \BibitemOpen
  \bibfield  {author} {\bibinfo {author} {\bibfnamefont {V.~V.}\ \bibnamefont
  {Kocharovsky}}, \bibinfo {author} {\bibfnamefont {V.~V.}\ \bibnamefont
  {Kocharovsky}}, \ and\ \bibinfo {author} {\bibfnamefont {M.~O.}\ \bibnamefont
  {Scully}},\ }\href@noop {} {\bibfield  {journal} {\bibinfo  {journal} {Phys.
  Rev. A}\ }\textbf {\bibinfo {volume} {61}},\ \bibinfo {pages} {053606}
  (\bibinfo {year} {2000})}\BibitemShut {NoStop}%
\bibitem [{\citenamefont {Bagnato}\ \emph {et~al.}(1987)\citenamefont
  {Bagnato}, \citenamefont {Pritchard},\ and\ \citenamefont
  {Kleppner}}]{Bagnato}%
  \BibitemOpen
  \bibfield  {author} {\bibinfo {author} {\bibfnamefont {V.}~\bibnamefont
  {Bagnato}}, \bibinfo {author} {\bibfnamefont {D.~E.}\ \bibnamefont
  {Pritchard}}, \ and\ \bibinfo {author} {\bibfnamefont {D.}~\bibnamefont
  {Kleppner}},\ }\href@noop {} {\bibfield  {journal} {\bibinfo  {journal}
  {Phys. Rev. A}\ }\textbf {\bibinfo {volume} {35}},\ \bibinfo {pages} {4354}
  (\bibinfo {year} {1987})}\BibitemShut {NoStop}%
\bibitem [{\citenamefont {Fleischhauer}\ and\ \citenamefont
  {Lukin}(2002)}]{PhysRevA.65.022314}%
  \BibitemOpen
  \bibfield  {author} {\bibinfo {author} {\bibfnamefont {M.}~\bibnamefont
  {Fleischhauer}}\ and\ \bibinfo {author} {\bibfnamefont {M.~D.}\ \bibnamefont
  {Lukin}},\ }\href {\doibase 10.1103/PhysRevA.65.022314} {\bibfield  {journal}
  {\bibinfo  {journal} {Phys. Rev. A}\ }\textbf {\bibinfo {volume} {65}},\
  \bibinfo {pages} {022314} (\bibinfo {year} {2002})}\BibitemShut {NoStop}%
\bibitem [{\citenamefont {Vadeiko}\ \emph {et~al.}(2005)\citenamefont
  {Vadeiko}, \citenamefont {Prokhorov}, \citenamefont {Rybin},\ and\
  \citenamefont {Arakelyan}}]{PhysRevA.72.013804}%
  \BibitemOpen
  \bibfield  {author} {\bibinfo {author} {\bibfnamefont {I.}~\bibnamefont
  {Vadeiko}}, \bibinfo {author} {\bibfnamefont {A.~V.}\ \bibnamefont
  {Prokhorov}}, \bibinfo {author} {\bibfnamefont {A.~V.}\ \bibnamefont
  {Rybin}}, \ and\ \bibinfo {author} {\bibfnamefont {S.~M.}\ \bibnamefont
  {Arakelyan}},\ }\href {\doibase 10.1103/PhysRevA.72.013804} {\bibfield
  {journal} {\bibinfo  {journal} {Phys. Rev. A}\ }\textbf {\bibinfo {volume}
  {72}},\ \bibinfo {pages} {013804} (\bibinfo {year} {2005})}\BibitemShut
  {NoStop}%
\bibitem [{\citenamefont {Demeter}\ \emph {et~al.}(2007)\citenamefont
  {Demeter}, \citenamefont {Dzsotjan},\ and\ \citenamefont
  {Djotyan}}]{PhysRevA.76.023827}%
  \BibitemOpen
  \bibfield  {author} {\bibinfo {author} {\bibfnamefont {G.}~\bibnamefont
  {Demeter}}, \bibinfo {author} {\bibfnamefont {D.}~\bibnamefont {Dzsotjan}}, \
  and\ \bibinfo {author} {\bibfnamefont {G.~P.}\ \bibnamefont {Djotyan}},\
  }\href {\doibase 10.1103/PhysRevA.76.023827} {\bibfield  {journal} {\bibinfo
  {journal} {Phys. Rev. A}\ }\textbf {\bibinfo {volume} {76}},\ \bibinfo
  {pages} {023827} (\bibinfo {year} {2007})}\BibitemShut {NoStop}%
\bibitem [{\citenamefont {Fleischhaker}\ \emph {et~al.}(2010)\citenamefont
  {Fleischhaker}, \citenamefont {Dey},\ and\ \citenamefont
  {Evers}}]{PhysRevA.82.013815}%
  \BibitemOpen
  \bibfield  {author} {\bibinfo {author} {\bibfnamefont {R.}~\bibnamefont
  {Fleischhaker}}, \bibinfo {author} {\bibfnamefont {T.~N.}\ \bibnamefont
  {Dey}}, \ and\ \bibinfo {author} {\bibfnamefont {J.}~\bibnamefont {Evers}},\
  }\href {\doibase 10.1103/PhysRevA.82.013815} {\bibfield  {journal} {\bibinfo
  {journal} {Phys. Rev. A}\ }\textbf {\bibinfo {volume} {82}},\ \bibinfo
  {pages} {013815} (\bibinfo {year} {2010})}\BibitemShut {NoStop}%
\bibitem [{\citenamefont {Agrawal}(2001)}]{Agrawal:book:2001}%
  \BibitemOpen
  \bibfield  {author} {\bibinfo {author} {\bibfnamefont {G.}~\bibnamefont
  {Agrawal}},\ }\href@noop {} {\emph {\bibinfo {title} {Applications of
  Nonlinear Fiber Optics (Optics and Photonics)}}}\ (\bibinfo  {publisher}
  {Academic Press},\ \bibinfo {year} {2001})\BibitemShut {NoStop}%
\bibitem [{\citenamefont {Akhmediev}\ and\ \citenamefont
  {Ankiewicz}(2008)}]{Lect}%
  \BibitemOpen
  \bibfield  {author} {\bibinfo {author} {\bibfnamefont {N.}~\bibnamefont
  {Akhmediev}}\ and\ \bibinfo {author} {\bibfnamefont {A.}~\bibnamefont
  {Ankiewicz}},\ }\href@noop {} {\emph {\bibinfo {title} {Dissipative Solitons:
  From Optics to Biology and Medicine (Lecture Notes in Physics)}}}\ (\bibinfo
  {publisher} {Springer},\ \bibinfo {year} {2008})\BibitemShut {NoStop}%
\bibitem [{\citenamefont {Mihalache}\ \emph {et~al.}(2007)\citenamefont
  {Mihalache}, \citenamefont {Mazilu}, \citenamefont {Lederer}, \citenamefont
  {Leblond},\ and\ \citenamefont {Malomed}}]{PhysRevA.76.045803}%
  \BibitemOpen
  \bibfield  {author} {\bibinfo {author} {\bibfnamefont {D.}~\bibnamefont
  {Mihalache}}, \bibinfo {author} {\bibfnamefont {D.}~\bibnamefont {Mazilu}},
  \bibinfo {author} {\bibfnamefont {F.}~\bibnamefont {Lederer}}, \bibinfo
  {author} {\bibfnamefont {H.}~\bibnamefont {Leblond}}, \ and\ \bibinfo
  {author} {\bibfnamefont {B.~A.}\ \bibnamefont {Malomed}},\ }\href {\doibase
  10.1103/PhysRevA.76.045803} {\bibfield  {journal} {\bibinfo  {journal} {Phys.
  Rev. A}\ }\textbf {\bibinfo {volume} {76}},\ \bibinfo {pages} {045803}
  (\bibinfo {year} {2007})}\BibitemShut {NoStop}%
\bibitem [{\citenamefont {Mihalache}\ \emph {et~al.}(2010)\citenamefont
  {Mihalache}, \citenamefont {Mazilu}, \citenamefont {Skarka}, \citenamefont
  {Malomed}, \citenamefont {Leblond}, \citenamefont
  {Aleksi\ifmmode~\acute{c}\else \'{c}\fi{}},\ and\ \citenamefont
  {Lederer}}]{PhysRevA.82.023813}%
  \BibitemOpen
  \bibfield  {author} {\bibinfo {author} {\bibfnamefont {D.}~\bibnamefont
  {Mihalache}}, \bibinfo {author} {\bibfnamefont {D.}~\bibnamefont {Mazilu}},
  \bibinfo {author} {\bibfnamefont {V.}~\bibnamefont {Skarka}}, \bibinfo
  {author} {\bibfnamefont {B.~A.}\ \bibnamefont {Malomed}}, \bibinfo {author}
  {\bibfnamefont {H.}~\bibnamefont {Leblond}}, \bibinfo {author} {\bibfnamefont
  {N.~B.}\ \bibnamefont {Aleksi\ifmmode~\acute{c}\else \'{c}\fi{}}}, \ and\
  \bibinfo {author} {\bibfnamefont {F.}~\bibnamefont {Lederer}},\ }\href
  {\doibase 10.1103/PhysRevA.82.023813} {\bibfield  {journal} {\bibinfo
  {journal} {Phys. Rev. A}\ }\textbf {\bibinfo {volume} {82}},\ \bibinfo
  {pages} {023813} (\bibinfo {year} {2010})}\BibitemShut {NoStop}%
\bibitem [{\citenamefont {Fedorov}\ \emph {et~al.}(2003)\citenamefont
  {Fedorov}, \citenamefont {Rosanov}, \citenamefont {Shatsev}, \citenamefont
  {Veretenov},\ and\ \citenamefont {Vladimirov}}]{Fedorov:2003}%
  \BibitemOpen
  \bibfield  {author} {\bibinfo {author} {\bibfnamefont {S.}~\bibnamefont
  {Fedorov}}, \bibinfo {author} {\bibfnamefont {N.}~\bibnamefont {Rosanov}},
  \bibinfo {author} {\bibfnamefont {A.}~\bibnamefont {Shatsev}}, \bibinfo
  {author} {\bibfnamefont {N.}~\bibnamefont {Veretenov}}, \ and\ \bibinfo
  {author} {\bibfnamefont {A.}~\bibnamefont {Vladimirov}},\ }\href {\doibase
  10.1109/JQE.2002.807212} {\bibfield  {journal} {\bibinfo  {journal} {Quantum
  Electronics, IEEE Journal of}\ }\textbf {\bibinfo {volume} {39}},\ \bibinfo
  {pages} {197 } (\bibinfo {year} {2003})}\BibitemShut {NoStop}%
\bibitem [{\citenamefont {Pontryagin}(1974)}]{Diff}%
  \BibitemOpen
  \bibfield  {author} {\bibinfo {author} {\bibfnamefont {L.}~\bibnamefont
  {Pontryagin}},\ }\href@noop {} {\emph {\bibinfo {title} {Obyknovennie
  differencialnie uravneniya}}}\ (\bibinfo  {publisher} {Nauka},\ \bibinfo
  {year} {1974})\BibitemShut {NoStop}%
\bibitem [{\citenamefont {Skarka}\ \emph
  {et~al.}(2010{\natexlab{b}})\citenamefont {Skarka}, \citenamefont
  {Aleksi\ifmmode~\acute{c}\else \'{c}\fi{}}, \citenamefont {Derbazi},\ and\
  \citenamefont {Berezhiani}}]{PhysRevB.81.035202}%
  \BibitemOpen
  \bibfield  {author} {\bibinfo {author} {\bibfnamefont {V.}~\bibnamefont
  {Skarka}}, \bibinfo {author} {\bibfnamefont {N.~B.}\ \bibnamefont
  {Aleksi\ifmmode~\acute{c}\else \'{c}\fi{}}}, \bibinfo {author} {\bibfnamefont
  {M.}~\bibnamefont {Derbazi}}, \ and\ \bibinfo {author} {\bibfnamefont
  {V.~I.}\ \bibnamefont {Berezhiani}},\ }\href {\doibase
  10.1103/PhysRevB.81.035202} {\bibfield  {journal} {\bibinfo  {journal} {Phys.
  Rev. B}\ }\textbf {\bibinfo {volume} {81}},\ \bibinfo {pages} {035202}
  (\bibinfo {year} {2010}{\natexlab{b}})}\BibitemShut {NoStop}%
\bibitem [{\citenamefont {Rozanov}\ \emph {et~al.}(2003)\citenamefont
  {Rozanov}, \citenamefont {Fedorov},\ and\ \citenamefont
  {Shatsev}}]{Rozanov_OS:2003}%
  \BibitemOpen
  \bibfield  {author} {\bibinfo {author} {\bibfnamefont {N.}~\bibnamefont
  {Rozanov}}, \bibinfo {author} {\bibfnamefont {S.}~\bibnamefont {Fedorov}}, \
  and\ \bibinfo {author} {\bibfnamefont {A.}~\bibnamefont {Shatsev}},\ }\href
  {\doibase 10.1134/1.1635463} {\bibfield  {journal} {\bibinfo  {journal}
  {Optics and Spectroscopy}\ }\textbf {\bibinfo {volume} {95}},\ \bibinfo
  {pages} {843} (\bibinfo {year} {2003})}\BibitemShut {NoStop}%
\bibitem [{\citenamefont {Johnson}\ and\ \citenamefont
  {Marburger}(1971)}]{PhysRevA.4.1175}%
  \BibitemOpen
  \bibfield  {author} {\bibinfo {author} {\bibfnamefont {R.~V.}\ \bibnamefont
  {Johnson}}\ and\ \bibinfo {author} {\bibfnamefont {J.~H.}\ \bibnamefont
  {Marburger}},\ }\href {\doibase 10.1103/PhysRevA.4.1175} {\bibfield
  {journal} {\bibinfo  {journal} {Phys. Rev. A}\ }\textbf {\bibinfo {volume}
  {4}},\ \bibinfo {pages} {1175} (\bibinfo {year} {1971})}\BibitemShut
  {NoStop}%
\bibitem [{\citenamefont {Pitaevskii}\ and\ \citenamefont
  {Lifshitz}(1980)}]{Stat2}%
  \BibitemOpen
  \bibfield  {author} {\bibinfo {author} {\bibfnamefont {L.~P.}\ \bibnamefont
  {Pitaevskii}}\ and\ \bibinfo {author} {\bibfnamefont {E.~M.}\ \bibnamefont
  {Lifshitz}},\ }\href@noop {} {\emph {\bibinfo {title} {Statistical Physics,
  Part 2: Volume 9 (Course of Theoretical Physics Vol. 9)}}}\ (\bibinfo
  {publisher} {Butterworth-Heinemann},\ \bibinfo {year} {1980})\BibitemShut
  {NoStop}%
\bibitem [{\citenamefont {Politzer}(1996)}]{Politzer}%
  \BibitemOpen
  \bibfield  {author} {\bibinfo {author} {\bibfnamefont {H.~D.}\ \bibnamefont
  {Politzer}},\ }\href@noop {} {\bibfield  {journal} {\bibinfo  {journal}
  {Phys. Rev. A}\ }\textbf {\bibinfo {volume} {54}},\ \bibinfo {pages} {1050}
  (\bibinfo {year} {1996})}\BibitemShut {NoStop}%
\bibitem [{\citenamefont {Balykin}\ \emph {et~al.}(1996)\citenamefont
  {Balykin}, \citenamefont {Laryushin}, \citenamefont {Subbotin},\ and\
  \citenamefont {Letokhov}}]{Balykin:1996}%
  \BibitemOpen
  \bibfield  {author} {\bibinfo {author} {\bibfnamefont {V.}~\bibnamefont
  {Balykin}}, \bibinfo {author} {\bibfnamefont {D.}~\bibnamefont {Laryushin}},
  \bibinfo {author} {\bibfnamefont {M.}~\bibnamefont {Subbotin}}, \ and\
  \bibinfo {author} {\bibfnamefont {V.}~\bibnamefont {Letokhov}},\ }\href
  {\doibase 10.1134/1.567094} {\bibfield  {journal} {\bibinfo  {journal}
  {Journal of Experimental and Theoretical Physics Letters}\ }\textbf {\bibinfo
  {volume} {63}},\ \bibinfo {pages} {802} (\bibinfo {year} {1996})}\BibitemShut
  {NoStop}%
\bibitem [{\citenamefont {Petrov}\ \emph {et~al.}(2000)\citenamefont {Petrov},
  \citenamefont {Shlyapnikov},\ and\ \citenamefont {Walraven}}]{Petrov}%
  \BibitemOpen
  \bibfield  {author} {\bibinfo {author} {\bibfnamefont {D.~S.}\ \bibnamefont
  {Petrov}}, \bibinfo {author} {\bibfnamefont {G.}~\bibnamefont {Shlyapnikov}},
  \ and\ \bibinfo {author} {\bibfnamefont {J.~T.~M.}\ \bibnamefont
  {Walraven}},\ }\href@noop {} {\bibfield  {journal} {\bibinfo  {journal}
  {Phys. Rev. Lett.}\ }\textbf {\bibinfo {volume} {85}},\ \bibinfo {pages}
  {3745} (\bibinfo {year} {2000})}\BibitemShut {NoStop}%
\end{thebibliography}%

\end{document}